\documentclass{emulateapj}
\usepackage{lscape}
\usepackage{apjfonts}
\usepackage{latexsym}
\shorttitle{RR Lyrae variables in M32 and M31} 
\shortauthors{Fiorentino et al.}
\begin{document}
\title{RR Lyrae variables in M32 and the disk of M31\footnotemark[1]}
\footnotetext[1]{Based on observations made with the NASA/ESA Hubble
  Space Telescope, obtained at the Space Telescope Science Institute,
  which is operated by the Association of Universities for Research in
  Astronomy, Inc., under NASA contract NAS 5-26555. These observations
  are associated with GO proposal 10572.}  

\author{Giuliana Fiorentino, Antonela Monachesi, Scott C. Trager} 
\affil{Kapteyn Astronomical Institute, Postbus 800, 9700 AV Groningen, The
  Netherlands} 
\email{fiorentino@astro.rug.nl} 
\author{Tod R. Lauer, Abhijit Saha, Kenneth J. Mighell} 
\affil{National Optical Astronomy Observatory\footnotemark[2], P.O.~Box 26732, Tucson, AZ, 85726, USA}
\footnotetext[2]{The National Optical Astronomy Observatory is
  operated by AURA, Inc., under cooperative agreement with the
  National Science Foundation.}  
\author{Wendy Freedman, Alan Dressler} 
\affil{The Observatories of the Carnegie Institution of Washington, 813 Santa Barbara Street, Pasadena, CA, 91101, USA}
\author{Carl Grillmair} 
\affil{Spitzer Science Center, 1200 E. California Blvd., Pasadena, CA 91125, USA} 
\and 
\author{Eline Tolstoy} \affil{Kapteyn Astronomical Institute, Postbus 800, 9700 AV Groningen, The Netherlands}

\begin{abstract}
  We observed two fields near M32 with the \textsl{Advanced Camera for
    Surveys/High Resolution Channel} (ACS/HRC) on board the
    \textsl{Hubble Space Telescope} (HST). The main field, F1, is
    1\farcm8 from the center of M32; the second field, F2, constrains
    the M31 background, and is 5\farcm4 distant.  Each field was
    observed for 16-orbits in each of the $F435W$ (narrow $B$) and
    $F555W$ (narrow $V$) filters.  The duration of the observations
    allowed RR Lyrae stars and other short-period variables to be
    detected.  A population of RR Lyrae stars determined to belong to
    M32 would prove the existence of an ancient population in that
    galaxy, a subject of some debate.

  We detected 17 RR Lyrae variables in F1 and 14 in F2.  A $1\sigma$
  upper limit of 6 RR Lyrae variables belonging to M32 is inferred
  from these two fields alone.  Use of our two ACS/WFC parallel fields
  provides better constraints on the M31 background, however, and
  implies that $7_{-3}^{+4}$ (68 \% confidence interval) RR Lyrae
  variables in F1 belong to M32.  We have therefore found evidence for
  an ancient population in M32.  It seems to be nearly
  indistinguishable from the ancient population of M31.  The RR Lyrae
  stars in the F1 and F2 fields have indistinguishable mean $V$-band
  magnitudes, mean periods, distributions in the Bailey diagram and
  ratios of RRc to RR$_{\mathrm{total}}$ types.  However, the color
  distributions in the two fields are different, with a population of
  red RRab variables in F1 not seen in F2.  We suggest that these
  might be identified with the detected M32 RR Lyrae population, but
  the small number of stars rules out a definitive claim.
\end{abstract}


\section{Introduction}
Messier 32 (M32) is the only elliptical galaxy close enough to
possibly allow direct observation of its stars down to the
main-sequence turn-off (MSTO).  It is a vital laboratory for
deciphering the stellar populations of all other elliptical galaxies,
which can only be studied by the spectra of their integrated light,
given their greater distances. Major questions about M32's star
formation history remain unanswered. M32 appears to have had one or
more relatively recent episodes of star formation \citep[within the
last 3 Gyr: e.g.,][]{OConnell80,Rose85,Rose94,G93,T00b,Coelho09},
which also appears to be true for many elliptical galaxies
\citep[e.g.,][]{G93,T00b,TMBO05}.  These conclusions rest on
painstaking and controversial spectral analysis of their integrated
light. In contrast, the most direct information about a stellar
population comes from applying stellar evolution theory to
color--magnitude diagrams (CMDs).  Little however is known about M32's
ancient population \citep[see, e.g.,][]{Brown00,Coelho09}.

With our \textsl{Advanced Camera for Surveys/High Resolution Channel}
(ACS/HRC) data (Cycle 14, Program GO-10572, PI: T. Lauer) we have
obtained the deepest CMD of M32 to date.  A comprehensive analysis of
this CMD is discussed in a companion paper \citet[hereafter M09]{M09}
and we refer to it for further details. However here we want to stress
that, due to the severe crowding in our fields, even with the high
spatial resolution of HRC it is not possible to reach the MSTO with
sufficient precision to claim the presence of a very old population.

RR Lyrae variables are low-mass stars burning He in their cores. They
are excellent tracers of ancient stellar populations, completely
independent of the MSTO, and knowledge of their properties provides
important information on their parent stellar populations. Because
they are located on the horizontal branch (HB) in a CMD, they are at
least 3 magnitudes brighter than MSTO dwarfs and therefore detectable
to relatively large distances. RR Lyrae are also very easy to
characterize, with ab-type RR Lyrae (RRab) pulsating in the
fundamental mode, rising rapidly to maximum light and slowly declining
to minimum light, and c-type RR Lyrae (RRc) pulsating in the first
harmonic mode, with their luminosities varying roughly sinusoidally.
Most importantly for our purpose, the mere presence of RR Lyrae stars
among a population of stars suggests an ancient origin, as ages older
than $\sim 10$ Gyr are required to produce RR Lyrae variables. Thus
the detection of RR Lyrae stars in M32 is presently the only way to
confirm the existence of an ancient stellar population in this galaxy.

\citet{AlonsoGarcia04} were the first to attempt to directly detect RR
Lyrae stars in fields near M32. They imaged a field $\sim3\farcm5$
with WFPC2 to the east of M32 and compared it with a control field
well away from M32 that should sample the M31 field stars. They
identified $12\pm8$ variable stars claimed to be RR Lyrae stars
belonging to M32 and therefore suggested that M32 possesses a
population that is older than $\sim10$ Gyr.  They were however unable
to classify these RR Lyrae variables and could not derive periods and
amplitudes for them. 

Very recently, \citet[hereafter S09]{s09} used ACS/WFC parallel
imaging from our present data set to find RR Lyrae variables in two
fields close to M32.  They found 681 RR Lyrae variables, with
excellent photometric and temporal completeness
(Sec.~\ref{sec:detection}).  These RR Lyrae stars were roughly equally
distributed between the two fields, with the same mean average
magnitudes, metallicities, and Oosterhoff types in each field.  It was
therefore impossible for them to separate the variables into M31 and
M32 populations.  \emph{It is still therefore an open question as to
  the precise nature or even presence of RR Lyrae variables in M32.}

In this paper, we present newly-detected RR Lyrae variables observed
with ACS/HRC and also a detailed analysis of the fields near M32 where
RR Lyrae stars have been found with HST. The paper is organized as
follows.  In Section 2 we describe our observations and the data
reduction we performed. We move on to describe the technique used to
identify and characterize the RR Lyrae variable stars in Section 3,
where we present their periods and light curves.  In Section 4 we show
that we have clearly detected RR Lyrae variables in M32, as long as we
include the results from our ACS/WFC parallel fields. In Section 5 we
discuss the properties of the RR Lyrae stars, such as the location of
their instability strip, reddenings, mean periods and Oosterhoff
types, as well as pulsational relations such as
period--metallicity--amplitude.  In this section we also derive
estimates of the distance moduli to and metallicities of our
fields. We summarize our findings and present our final conclusions in
Section 6.

\section{Observations and Data reduction}

\begin{deluxetable*}{lllccr}
 \tabletypesize{\scriptsize}    
\tablecaption{Log of observations\label{table:data}} 
\tablewidth{0pt}
\tablehead{\colhead{Field}&\colhead{$\alpha_{J2000.0}$}&
\colhead{$\delta_{J2000.0}$}&\colhead{Filter}&\colhead{Exposure time}&
\colhead{Date}} 
\startdata
F1&00 42 47.63&+40 50 27.4&$F435W$ &16$\times$1279 +
16$\times$1320&20--22 Sep 2005\\
F1&00 42 47.63&+40 50 27.4&$F555W$&16$\times$1279 +
16$\times$1320&22--24 Sep 2005\\ 
F2&00 43 \phn7.89&+40 54 14.5&$F435W$&16$\times$1279 +
16$\times$1320&6--8 Feb 2006\\ 
F2&00 43 \phn7.89&+40 54 14.5&$F555W$&16$\times$1279 +
16$\times$1320&9--12 Feb 2006
\enddata
\end{deluxetable*}


\subsection{Field Selection, Observational Strategy and Data
  Reduction}

We obtained deep $B$ and $V$-band imaging of two fields near M32 using
the ACS/HRC instrument on board HST during Cycle 14 (Program GO-10572,
PI: Lauer).  The primary goal of this program was to resolve the M32
MSTO.  The ACS $F435W$ ($B$) and $F555W$ ($V$) filters were selected
to optimize detection of MSTO stars over the redder and more luminous
stars of the giant branch.  M32 is very compact and is projected
against the M31 disk.  Thus the major challenge was to select a field
that represented the best compromise between the extreme crowding in
M32, which would drive the field to be placed as far away from the
center of the galaxy as possible, versus maximizing the contrast of
M32 against the M31 background populations, which would push the field
back towards the central, bright portions of M32.  Following these
constraints, the M32 HRC field (designated F1) was centered on a
location $110\arcsec$ south (the anti-M31 direction) of the M32
nucleus, roughly on the major axis of the galaxy.  The $V$-band
surface brightness of M32 near the center of the field is
$\mu_V\approx21.9$ \citep{kfc}.  M32 quickly becomes too crowded to
resolve faint stars at radii closer to the center, while the galaxy
rapidly falls below the M31 background at larger radii.

\begin{figure*}
\plotone{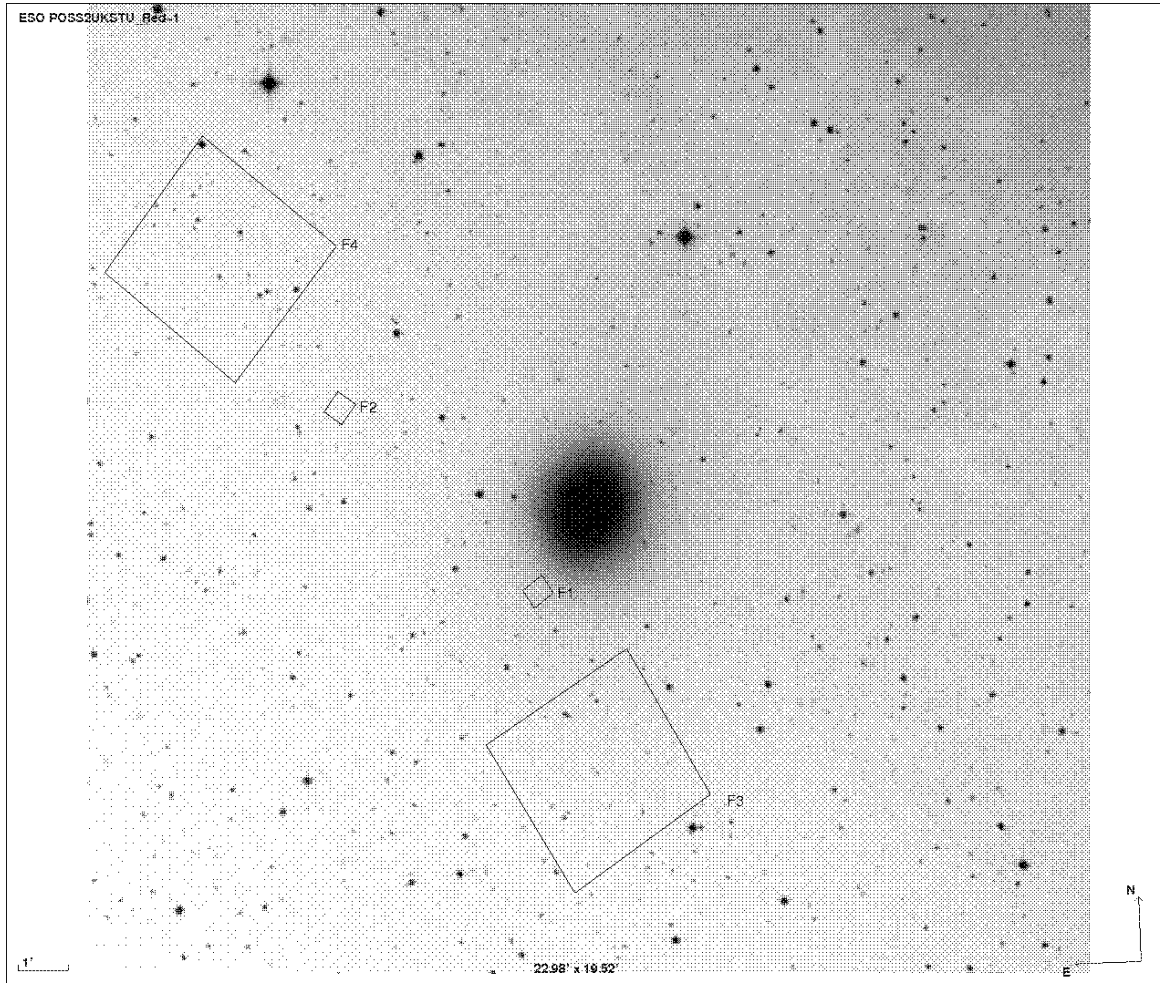}
\caption{Location of our pointings near M32 observed with ACS/HRC on
  board HST (small squares). Fields F1 and F2 are at distances of
  1\farcm8 and 5\farcm4 from the center of M32, respectively, and each
  of them covers a region of 0.25 arcmin$^2$ on the sky. Parallel
  fields, taken with ACS/WFC, F3 and F4 are at distances of 5\farcm3
  and 9\farcm2 from the center of M32, respectively, covering a
  region of 9 arcmin$^2$ on the sky.}
\label{pointings}
\end{figure*}

Even at the location of F1, M31 contributes $\sim1/3$ of the total
light, thus it was critical to obtain a background field, F2, at the
same isophotal level in M31 ($\mu_V\sim22.7$) to allow for the strong
M31 contamination to be subtracted from the analysis of the M32
stellar population.  F2 was located $327\arcsec$ from the M32 nucleus
at position angle $65^\circ.$ At this angular distance M32 has an
ellipticity $\epsilon\approx0.25$ \citep{choi}, and F2 is nearly
aligned with the M32 minor-axis.  Thus the implied semi-major axis of
the M32 isophote that passes though F2 is $435\arcsec$, significantly
larger than the nominal angular separation.  The estimated M32 surface
brightness at F2 is $\mu_V\approx27.5,$ based on a modest
extrapolation of the $B$-band surface photometry of \citet{choi} and
an assumed color of $B-V\approx0.9$.  The contribution of M32 to F2
thus falls by a factor of $\sim180$ relative to its surface brightness
at F1.  While one might have been tempted to move F2 even further away
from F1, it clearly serves as an adequate background at the location
selected, while uncertainties in the M31 background would increase at
larger angular offsets.  The locations of both the F1 and F2 fields
are shown in Figure~\ref{pointings}.

Detection of the MSTO required deep exposures at F1.  Accurate
treatment of the background required equally deep exposures to be
obtained in F2.  A summary of the observations is shown in
Table~\ref{table:data}; briefly, each field was observed for 16 orbits
in each of the $F435W$ and $F555W$ filters for a total program of 64
orbits.  While the detection of RR Lyrae variables was not the primary
goal of the program, execution of each filter/field combination in a
contiguous time span of 2--3 days was clearly well-suited to detect RR
Lyrae variables, which have periods ranging from 0.2--1 d.

At $B$ and $V$, the HRC undersamples the PSF, despite its
exceptionally fine pixel scale.  All of the images were obtained in a
$0.5\times0.5$ sub-pixel square dither pattern to obtain Nyquist
sampling in the complete data set.  In detail, the sub-pixel dither
pattern was executed across each pair of orbits, with each orbit split
into two sub-exposures.  The telescope was then offset by $0\farcs125$
steps between the orbit pairs in a ``square-spiral'' dither pattern to
minimize the effects of ``hot pixels,'' bad columns, and any other
fixed-defects in the CCD, on the photometry at any location.  The data
for each filter/field combination thus comprises 8 slightly different
pointings, with Nyquist-sampling obtained at each location.  In
practice, the dithers were extremely accurate, and Nyquist images
could readily be constructed using the algorithm of \citet{l99}.

Our optimal \emph{average} photometry has been obtained by using these
very deep, super-resolved images.  We have performed photometry by
using both the DAOPHOT II/ALLSTAR packages \citep{Stetson87,Stetson94}
and by first deconvolving those combined images with a reliable PSF
and then performing aperture photometry on the deconvolved
images. Both methods returned comparable results and allow us to
present the deepest CMD of M32 obtained so far.  This result is
analyzed in M09 and will not be discussed further here. However, in
what follows we will use these CMDs to show the location of RR Lyrae
stars.

In addition to the HRC images, parallel observations were obtained
with the ACS/WFC channel using the $F606W$ filter (broad $V$).  These
fields, designated F3 and F4, are also shown in
Figure~\ref{pointings}.  Notably, the telescope rolled by roughly
$180^\circ$ between the execution of the F1 and F2 observations, thus
F3, the parallel field associated with F1, and F4, the mate to F2,
bracket the F1 and F2 fields in angle.  By happenstance, the F3 and F4
fields also nearly fall on the same M31 isophote that encompasses the
F1 and F2 fields, thus the M31 background should be roughly similar in
all four fields.  It is also notable that F3 is positioned slightly
closer to the M32 nucleus than F2 ($317\arcsec$ versus $327\arcsec$),
but because it also falls along the M32 major rather than minor axis,
its associated M32 surface brightness is $\mu_V\sim25$, or a factor of
$\sim10\times$ more than the M32 contribution to F2.  Furthermore we
note that the parallel observations were exposed only at the same time
as the $F555W$ exposures in F1 and F2 and therefore cover only half of
the total time window of the primary exposures (we return to this
point in Sec.~\ref{sec:detection} below).

The parallel images have already been analyzed by S09.  They find 681
RR Lyrae variables stars, of which 324 are located in the field
closest to M32. Because only one filter was available for the parallel
observations, S09 did not have all the information needed to properly
disentangle the populations that belong to M31 and/or M32. In fact,
their detected RR Lyrae stars show the same mean average magnitude,
metallicity and Oosterhoff type, as we discuss in
Section~\ref{sec:properties}. With our primary observations we can
attempt to disentangle the two populations by using all the quantities
characterizing the class of RR Lyrae variables, such as mean weighted
magnitudes and colors in the Johnson-Cousins system, periods, and
amplitudes.

\subsection{Photometry of the RR Lyrae Variables}

The study of the presence of RR Lyrae stars is based on a detailed
analysis of the time series of our fields.  We analyzed each single
epoch image (32 per field and per filter) and not the combination of
all the images described above. Because of the intrinsic brightness of
the RR Lyrae ($V\sim25$ mag), we decided to perform PSF-fitting
photometry over all the fully calibrated data products (FLT) images
using the DOLPHOT package, a version of HSTphot \citep{Dolphin00}
modified for ACS images. Our choice has been justified by the short
time consumed to obtain high quality photometry at the RR Lyrae
magnitude level for our data set.  Following the DOLPHOT User's Guide,
we have performed the pre-processing steps \textsc{mask} and
\textsc{calcsky} routines before running DOLPHOT. This package
performs photometry simultaneously over all 64 images of each field,
returning a catalog of more than 20000 stars per field already
corrected for charge transport efficiency (CTE) and aperture
correction, following the suggestions by \citet{Sirianni05}. We use
this photometry to perform the analysis of variable stars.

\subsubsection{Photometric Completeness at the Horizontal Branch}

The 50\% completeness levels of fields F1 and F2 are at least 2 mag
deeper than the horizontal branch, as shown by the artificial star
tests (ASTs) in M09. The ASTs show that the completeness at the red
clump (i.e., the red horizontal branch) is 100\%.  The ASTs in M09 do
not properly populate the blue horizontal branch, as there are very
few stars in this region of the CMD, so we assume that the
completeness at the position of the blue horizontal branch is the same
as at the red clump.

\section{Searching for variable stars and their periods}
\label{sec:identification}
To identify variable sources in both fields, we used a code written by
one of us (AS) in the Interactive Data Language (IDL) whose
principles, based on the algorithm of \citet{LK65}, are discussed in
\citet{SH90}. This code was applied to the results from DOLPHOT
PSF-fitting photometry described in the previous section. Its output
gives us not only a list of candidate variable stars but also a good
initial estimate of their periods.  The method assumes that realistic
error estimates for each object at each epoch are available from the
photometry, which are first used to estimate a chi-square based
probability that any given object is a variable.  A list of candidates
is then chosen, and each candidate is tested for periodicity and
plausible light curves. The graphical interface of this program
clearly shows possible aliases and allows the user to examine the
light curves implied for each such alias. The final decision making is
done by the user. A refinement of the period for all the candidate
variables has been performed by using two other independent codes. We
used the Period Dispersion Minimization (PDM) algorithm in the IRAF
environment to confirm the found periodicity
\citep{Stellingwerf78}. Further refinement was then obtained by using
GRATIS (GRaphical Analyzer of TIme Series, developed by P. Montegriffo
at the Bologna Observatory; see \citealt{Clementini00} and references
therein for details), which permits us to fit Fourier series to the
magnitudes in each passband as a function of their phase.

\begin{deluxetable*}{@{}lcccccccccccccc}
\tablecaption{F1 RR Lyrae properties\label{table:RR_Lyraem32}} 
\tablewidth{0pt} 
\tablecolumns{15}
\tablehead{\colhead{Star id}& \colhead{RA} & \colhead{DEC} &
  \colhead{Period} &\colhead{H\tablenotemark{a}}&
  \colhead{Epoch\tablenotemark{b}}& \colhead{$F435W$} &
  \colhead{$F555W$} &\colhead{$\langle V\rangle$}
   & \colhead{$\langle B\rangle-\langle V\rangle$} & \colhead{$A_B$}
  &\colhead{$A_V$}&\colhead{res$B$\tablenotemark{c}}  
  &\colhead{res$V$\tablenotemark{c}}& \colhead{type} \\
  & \colhead{J2000} & \colhead{J2000} & \colhead{(days)} & & (JD) &
  \colhead{mag} & \colhead{mag}&\colhead{mag} & \colhead{mag}
  &\colhead{mag}  &\colhead{mag}& \colhead{mag} & \colhead{mag} &}
\startdata 
 1&00:42:48.435&+40:50:32.11&0.255&1&2453634.900&25.56&25.39&25.44&0.19&0.64&0.59&0.11&0.10&RRc\\   
 2&00:42:46.509&+40:50:30.19&0.285&1&2453633.060&25.68&25.45&25.50&0.24&0.77&0.56&0.12&0.11&RRc\\   
 3&00:42:46.563&+40:50:23.03&0.311&1&2453632.810&25.42&25.22&25.30&0.19&0.61&0.58&0.09&0.11&RRc\\   
 4&00:42:48.237&+40:50:12.85&0.317&1&2453632.900&25.59&25.43&25.47&0.20&0.61&0.42&0.12&0.10&RRc\\   
 5&00:42:48.384&+40:50:30.61&0.475&2&2453636.290&25.52&25.29&25.34&0.25&0.99&0.85&0.10&0.11&RRab\\  
 6&00:42:47.349&+40:50:41.91&0.486&3&2453634.450&25.93&25.56&25.60&0.41&1.09&0.82&0.15&0.15&RRab\\  
 7&00:42:48.074&+40:50:30.17&0.519&4&2453634.415&25.34&25.04&25.09&0.31&1.01&0.75&0.08&0.06&RRab\\  
 8&00:42:47.552&+40:50:30.41&0.521&3&2453634.560&25.68&25.41&25.46&0.28&0.93&0.76&0.09&0.10&RRab\\  
 9&00:42:47.087&+40:50:43.08&0.523&3&2453635.650&25.61&25.25&25.29&0.40&1.09&1.06&0.12&0.13&RRab\\  
10&00:42:47.034&+40:50:28.29&0.546&3&2453632.216&25.93&25.51&25.55&0.45&1.41&1.03&0.18&0.16&RRab\\  
11&00:42:47.462&+40:50:42.81&0.564&2&2453635.490&25.51&25.19&25.21&0.36&0.74&0.68&0.15&0.09&RRab\\  
12&00:42:48.737&+40:50:32.31&0.621&2&2453632.600&25.43&25.19&25.18&0.30&1.25&0.89&0.17&0.15&RRab\\  
13&00:42:47.014&+40:50:36.27&0.625&2&2453637.280&25.57&25.28&25.29&0.35&0.88&0.68&0.09&0.08&RRab\\  
14&00:42:46.411&+40:50:29.96&0.626&3&2453632.500&25.53&25.15&25.19&0.43&1.09&0.81&0.09&0.09&RRab\\  
15&00:42:47.554&+40:50:16.48&0.645&3&2453634.680&25.83&25.47&25.50&0.40&0.74&0.39&0.10&0.07&RRab\\  
16&00:42:46.075&+40:50:25.38&0.728&2&2453637.320&25.74&25.22&25.25&0.58&0.39&0.38&0.12&0.08&RRab\\  
17&00:42:46.638&+40:50:25.24&0.851&2&2453637.693&25.60&25.15&25.17&0.49&0.65&0.45&0.10&0.08&RRab\\  
\enddata
\tablenotetext{a}{Order of the Fourier series used to obtain
  the best fit} 
\tablenotetext{b}{Julian Date where each curve shows its
  maximum of light at phase $\phi=1$} 
\tablenotetext{c}{RMS deviation of the data points from the fitting
  model, in $B$- (res$B$) and $V$- (res$V$) bands respectively}
\end{deluxetable*}

\begin{deluxetable*}{@{}lcccccccccccccc}
\tablecaption{F2 RR Lyrae properties\label{table:RR_Lyraem31}} 
\tablewidth{0pt} 
\tablecolumns{15}
\tablehead{\colhead{Star id}& \colhead{RA} & \colhead{DEC} &
  \colhead{Period} &\colhead{H\tablenotemark{a}}&
  \colhead{Epoch\tablenotemark{b}}& \colhead{$F435W$} &
  \colhead{$F555W$} &\colhead{$\langle V\rangle$}
   & \colhead{$\langle B\rangle-\langle V\rangle$} & \colhead{$A_B$}
  &\colhead{$A_V$}&\colhead{res$B$\tablenotemark{c}}  
  &\colhead{res$V$\tablenotemark{c}}& \colhead{type} \\
  & \colhead{J2000} & \colhead{J2000} & \colhead{(days)} & & (JD) &
  \colhead{mag} & \colhead{mag}&\colhead{mag} & \colhead{mag}
  &\colhead{mag}  &\colhead{mag}& \colhead{mag} & \colhead{mag} &}
\startdata 
 1&00:43:07.766&+40:54:15.31&0.267&2&2453772.400&25.46&25.32&25.38&0.20&0.66&0.59&0.09&0.09&RRc\\    
 2&00:43:07.704&+40:54:23.18&0.287&2&2453774.700&25.47&25.36&25.41&0.18&0.70&0.55&0.10&0.07&RRc\\    
 3&00:43:08.518&+40:54:08.86&0.320&2&2453774.125&25.51&25.26&25.31&0.33&0.46&0.43&0.08&0.09&RRc\\    
 4&00:43:07.734&+40:54:27.17&0.326&2&2453774.310&25.51&25.36&25.41&0.24&0.57&0.47&0.10&0.09&RRc\\    
 5&00:43:07.874&+40:54:31.99&0.350&1&2453777.220&25.45&25.25&25.30&0.28&0.60&0.42&0.11&0.09&RRc\\    
 6&00:43:08.300&+40:54:21.88&0.383&2&2453771.830&25.25&25.08&25.14&0.23&0.63&0.49&0.07&0.07&RRc\\    
 7&00:43:07.671&+40:54:01.99&0.482&3&2453778.050&25.53&25.33&25.39&0.28&1.37&1.08&0.15&0.12&RRab\\   
 8&00:43:08.339&+40:54:23.36&0.502&3&2453774.324&25.64&25.32&25.37&0.39&1.19&1.05&0.12&0.14&RRab\\   
 9&00:43:07.594&+40:54:21.32&0.528&3&2453773.198&25.53&25.27&25.30&0.35&1.27&1.02&0.11&0.11&RRab\\   
10&00:43:07.424&+40:54:26.69&0.528&3&2453774.846&25.40&25.17&25.23&0.29&1.26&0.89&0.13&0.12&RRab\\   
11&00:43:07.727&+40:54:13.25&0.571&2&2453775.520&25.52&25.28&25.35&0.28&1.06&0.99&0.15&0.14&RRab\\   
12&00:43:08.107&+40:54:21.00&0.588&3&2453775.045&25.64&25.36&25.41&0.35&0.94&0.81&0.10&0.10&RRab\\   
13&00:43:07.977&+40:54:21.33&0.697&2&2453778.053&25.40&25.12&25.19&0.33&0.70&0.58&0.09&0.06&RRab\\   
14&00:43:07.798&+40:54:14.39&0.790&2&2453778.450&25.28&24.95&25.00&0.40&0.63&0.56&0.07&0.08&RRab\\   
\enddata 
\tablenotetext{a}{Order of the Fourier series used to obtain
  the best fit} 
\tablenotetext{b}{Julian Date where each curve shows its
  maximum of light at phase $\phi=1$} 
\tablenotetext{c}{RMS deviation of the data points from the fitting
  model, in $B$- (res$B$) and $V$- (res$V$) bands respectively}
\end{deluxetable*}

\begin{deluxetable}{@{}lccccr}
  \tabletypesize{\scriptsize} \tablecaption{Time-series magnitudes for
    RR Lyrae variables\label{table:serie}}
    \tablewidth{0pt} 
    \tablecolumns{6}
    \tablehead{ \colhead{Julian date}&\colhead{F435W}&\colhead{$B$}& \colhead{Julian date}&\colhead{F555W}&\colhead{$V$}\\
    \colhead{$-$2400000}&\colhead{mag}&\colhead{mag}&\colhead{$-$2400000}&\colhead{mag}&\colhead{mag}}  
    \startdata 
\cutinhead{F1 variable 1}\\
    53633.616&25.39$\pm$0.07&25.44&53635.480&25.42$\pm$0.07&25.48\\
    53633.631&25.22$\pm$0.06&25.28&53635.496&25.60$\pm$0.08&25.66\\
    53633.681&25.36$\pm$0.07&25.42&53635.545&25.72$\pm$0.08&25.78\\
    53633.697&25.69$\pm$0.09&25.74&53635.560&25.62$\pm$0.08&25.67\\
    53633.748&25.96$\pm$0.10&26.01&53635.612&25.44$\pm$0.06&25.50\\
    53633.764&25.89$\pm$0.10&25.93&53635.628&25.24$\pm$0.06&25.29\\
    53633.814&25.86$\pm$0.10&25.90&53635.678&25.25$\pm$0.06&25.31\\
    53633.830&25.50$\pm$0.07&25.55&53635.694&25.24$\pm$0.06&25.30\\
    53634.415&25.39$\pm$0.08&25.44&53636.213&25.17$\pm$0.06&25.22\\
    53634.430&25.37$\pm$0.08&25.42&53636.228&25.38$\pm$0.07&25.44\\
     \enddata  
     \tablecomments{The errors on HST VEGAMAG are the photometrical
       ones. The calibration onto the Johnson-Cousins $B$- and
       $V$-bands as well as their errors have been discussed in the
       text. This is a sample table showing the format of the Table;
       the complete table can be found in the online version of the
       journal.}
\end{deluxetable}

The magnitudes returned by DOLPHOT have already been calibrated onto
the HST VEGAMAG photometric system, but for the following analysis we
need to transform them onto the Johnson-Cousins (JC) system. We need
therefore to take into account the color variations in the periodic
cycles of the variable stars. We thus associate each phased epoch in
the $F435W$ filter with the corresponding best-fitting $F555W$ model
provided by GRATIS at that epoch, and vice versa. Finally, we apply
the \citet{Sirianni05} transformations from $F435W$ and $F555W$ to $B$
and $V$ for ACS/HRC, and we re-analyze the new JC time series with
GRATIS to improve the light curve models as well as the
previously-constrained periods. The order of the Fourier series used
to obtain the best fit and the epoch corresponding to maximum of the
light curve at phase $\phi=1$ are given in columns 5 and 6 in
Tables~\ref{table:RR_Lyraem32} and \ref{table:RR_Lyraem31}. This
procedure allows us to derive well-sampled and consistent light curves
in both filters. Proper periods, mean magnitudes weighted on the
proper light curve in both HST VEGAMAG and JC photometric systems,
colors, and amplitudes are given in Tables~\ref{table:RR_Lyraem32} and
\ref{table:RR_Lyraem31} for a total number of 31 \emph{bona fide} RR
Lyrae stars: 17 in F1 and 14 in F2. The time series photometry is
given in Table \ref{table:serie}.

\begin{figure*}
\plotone{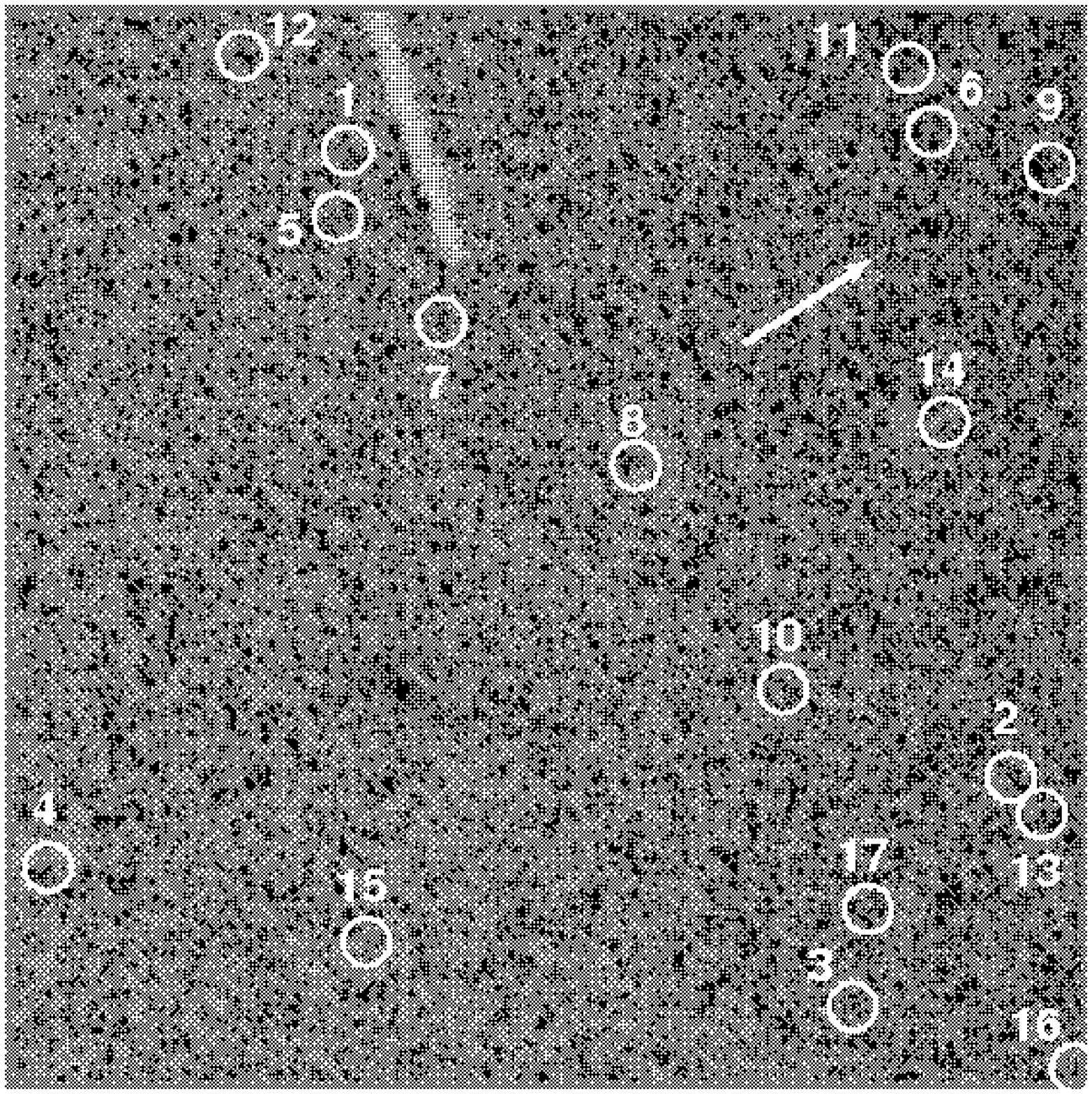}
\caption{Finding chart for RR Lyrae variables found in field
  F1. Numbers correspond to variables listed in
  Table~\ref{table:RR_Lyraem32}. The FoV size is 0.25 arcmin$^2$, as
  listed in column 9 of Table~\ref{table:grad}. We note that the RR
  Lyrae stars in F1 are slightly clustered along the edge closest to
  the center of M32 where the total stellar density is increasing; we
  return to this point in Section~\ref{sec:PIS}. The arrow points
  towards the center of M32.}
\label{FC-F1}
\end{figure*}

\begin{figure*}
\plotone{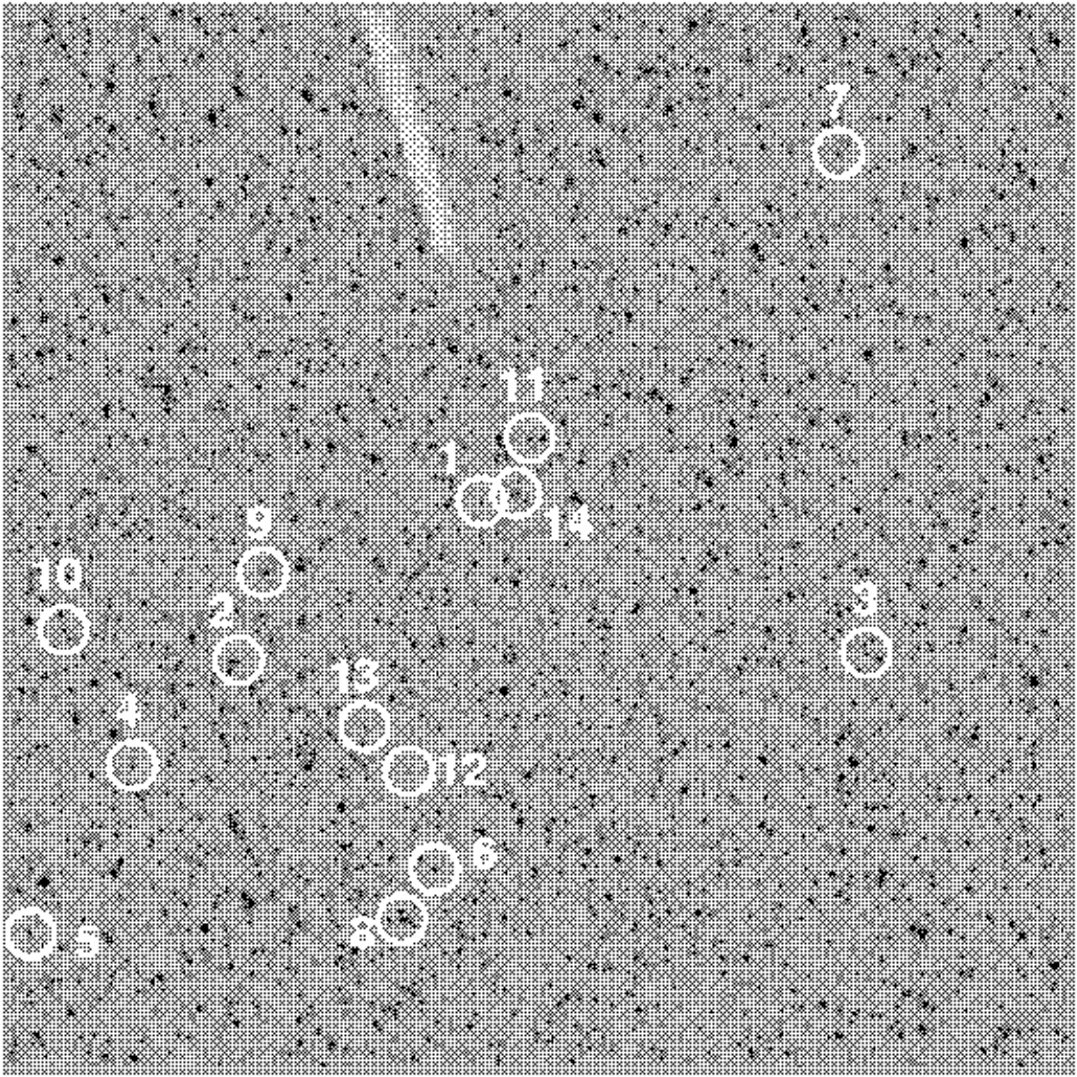}
\caption{Finding charts for RR Lyrae variables found in field F2.
  Numbers correspond to variables listed in
  Table~\ref{table:RR_Lyraem31}. The FoV size is 0.25 arcmin$^2$, as
  listed in column 9 of Table~\ref{table:grad}. The adopted
  intensity scale is the same as in Fig.~\ref{FC-F1} for a fair
  comparison.}
\label{FC-F2}
\end{figure*}

\begin{figure*}
\plottwo{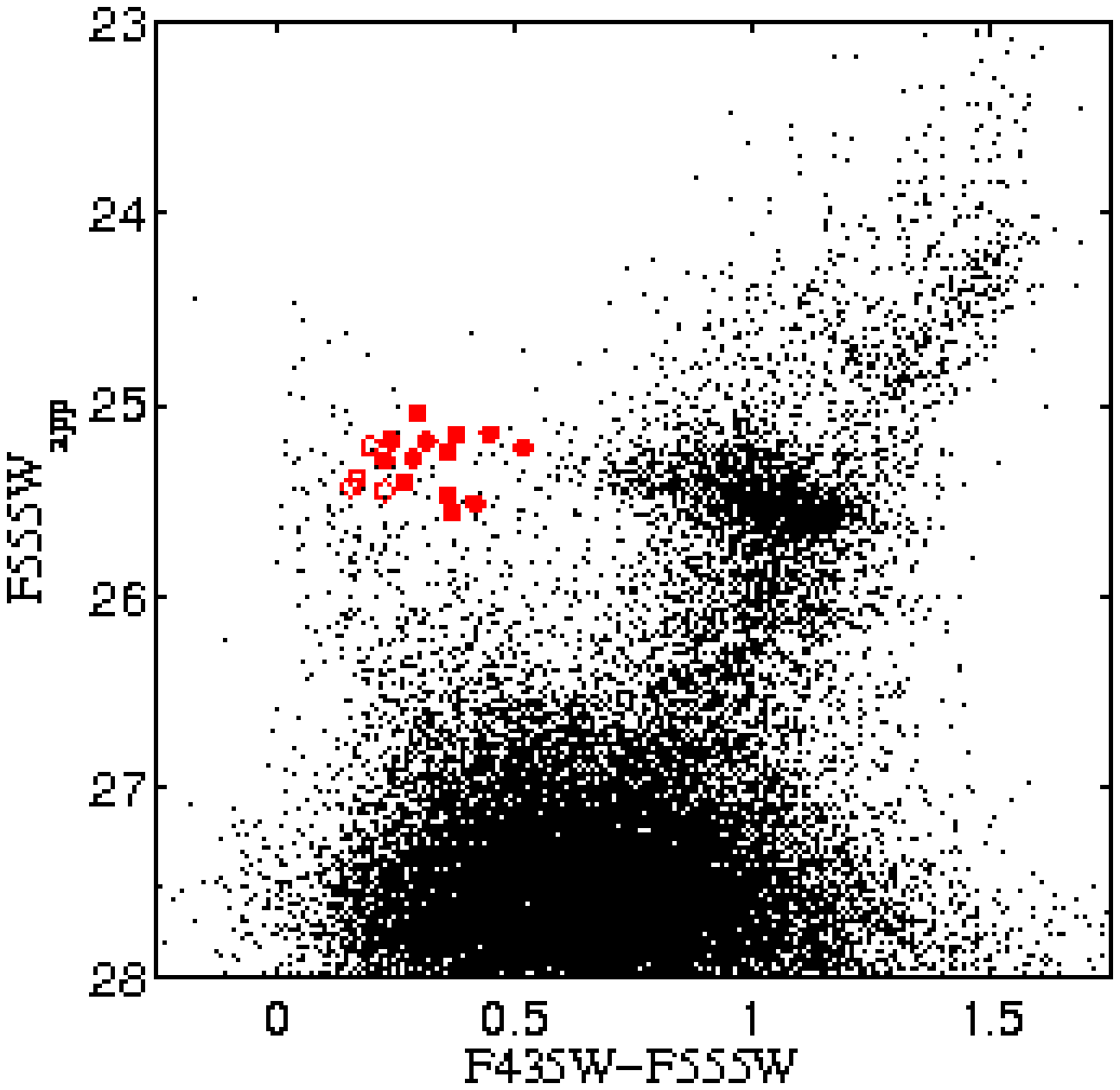}{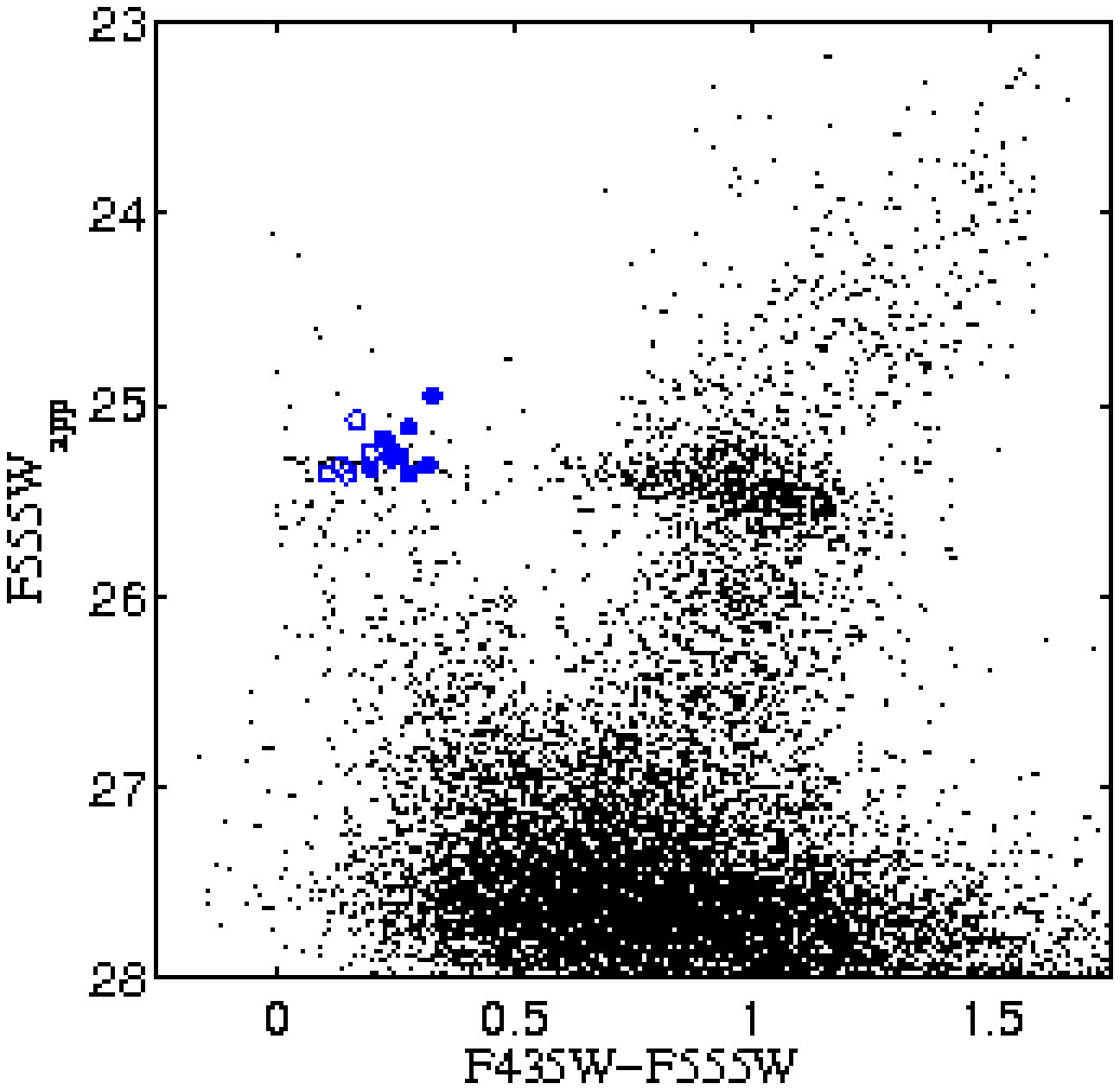}
\caption{The ($F435W-F555W$, $F555W$) CMDs calibrated onto the HST
  VEGAMAG photometric system for fields F1 (left) and F2 (right).  We
  show the location of the detected RR Lyrae variable
  stars. First-overtone (FO) and fundamental-mode (FU) pulsators are
  shown with empty and filled circles respectively.  These CMDs are
  presented and discussed in M09.}
\label{cmd-F1F2}
\end{figure*}

\begin{figure*}
\plottwo{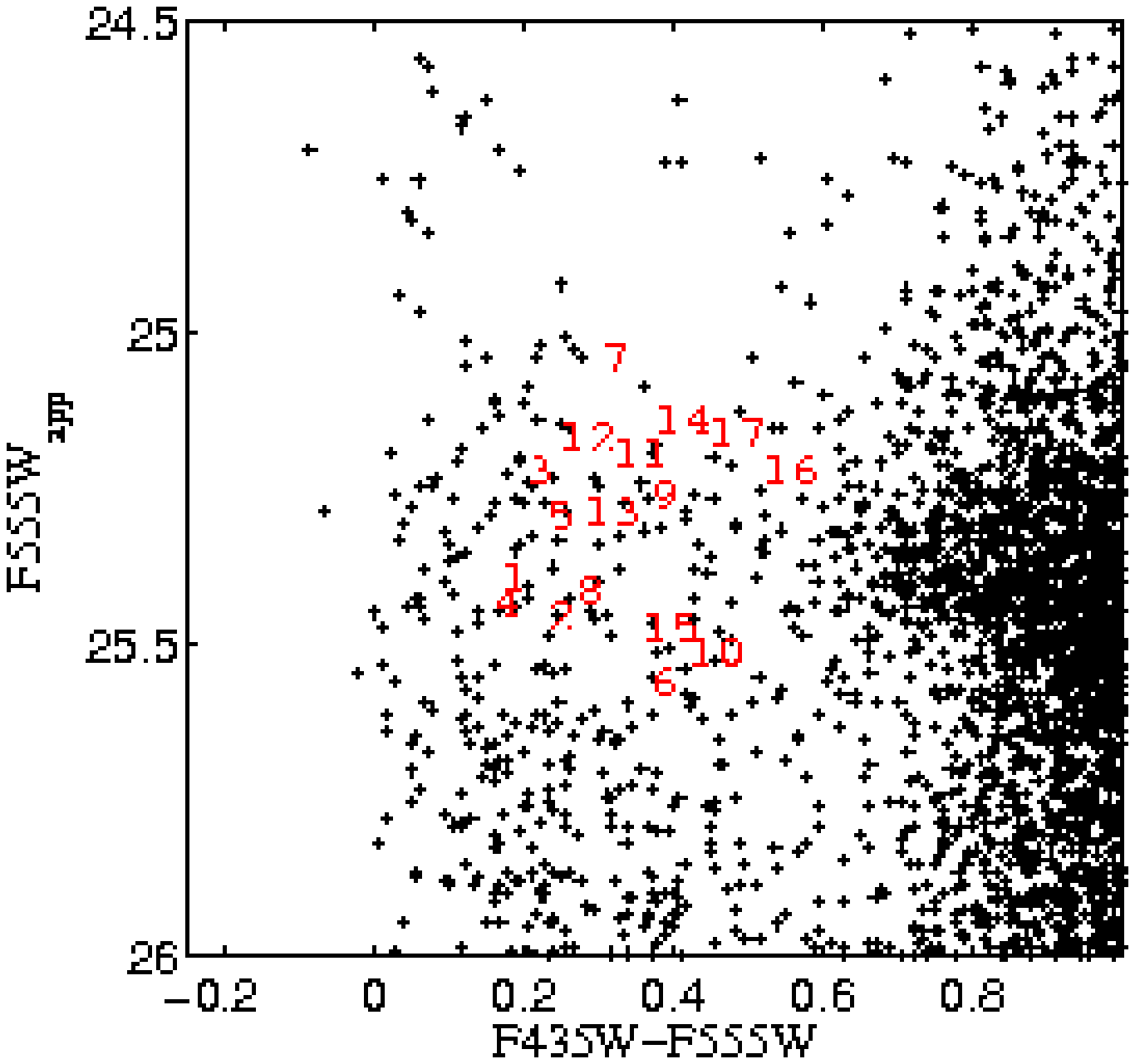}{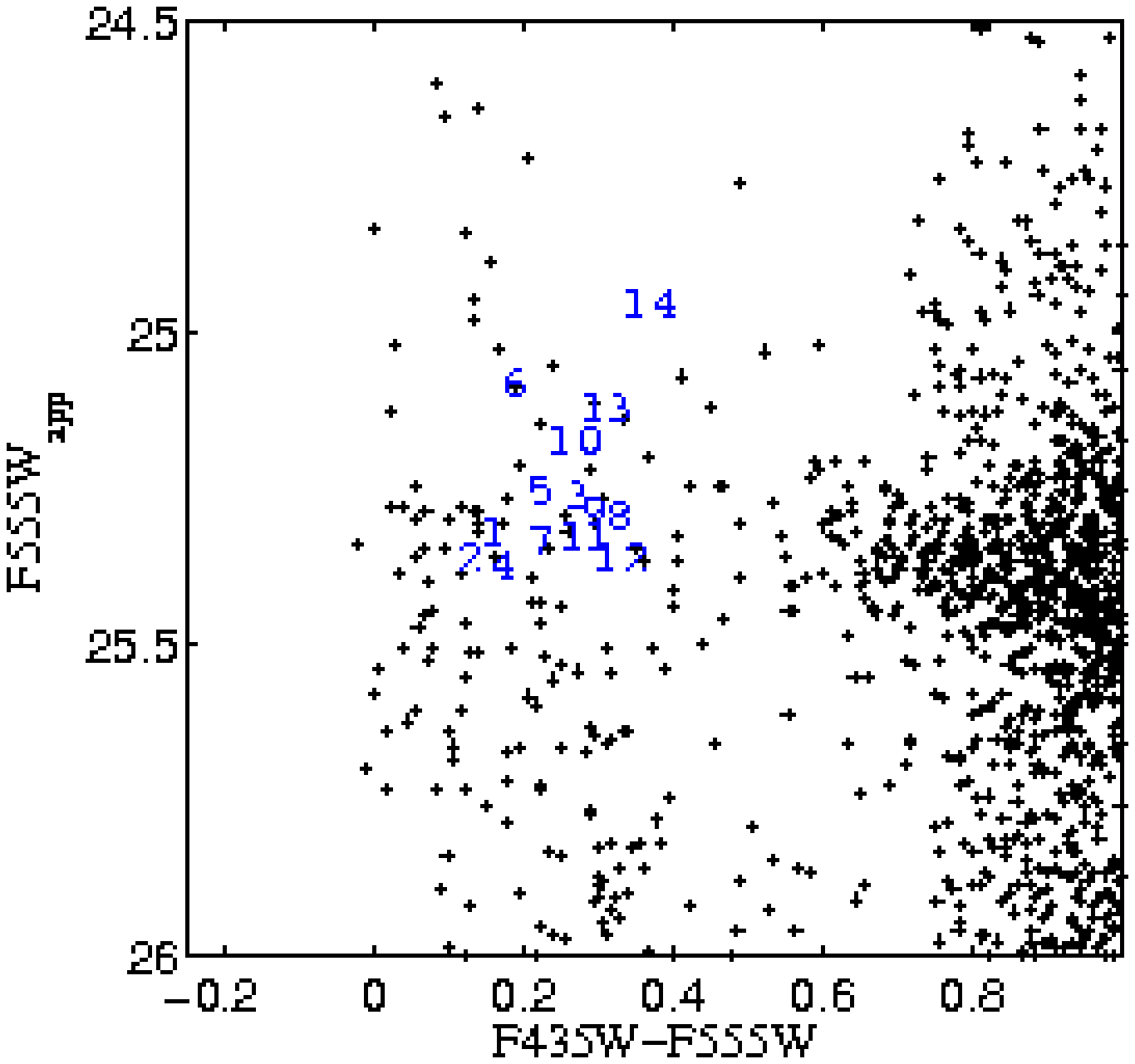}
\caption{As in Fig.~\ref{cmd-F1F2}, zoomed into the region of the
  detected RR Lyrae stars.  Left: F1.  Right: F2. Numbers correspond
  to variables listed in Tables~\ref{table:RR_Lyraem32} (left panel)
  and \ref{table:RR_Lyraem31} (right panel), respectively.}
\label{cmd-F1F2zoom}
\end{figure*}

Finding charts for the newly-detected RR Lyrae variables are shown in
Figures~\ref{FC-F1} and \ref{FC-F2} on the combined $F555W$ images of
F1 and F2, respectively. Their locations are shown in the CMDs from
M09 calibrated onto the HST VEGAMAG photometric system using the
average magnitudes as reported in Tables~\ref{table:RR_Lyraem32} and
\ref{table:RR_Lyraem31} (Figure~\ref{cmd-F1F2}). We cross-correlated
the RR Lyrae coordinates and magnitudes with the photometric catalog
from M09 to confirm the presence of our new RR Lyrae stars in the
average photometry. For all the RR Lyrae variables we found a star
with same coordinates and similar magnitude (see the zoomed-in CMDs in
Fig.~\ref{cmd-F1F2zoom}).

\begin{figure*}
\plotone{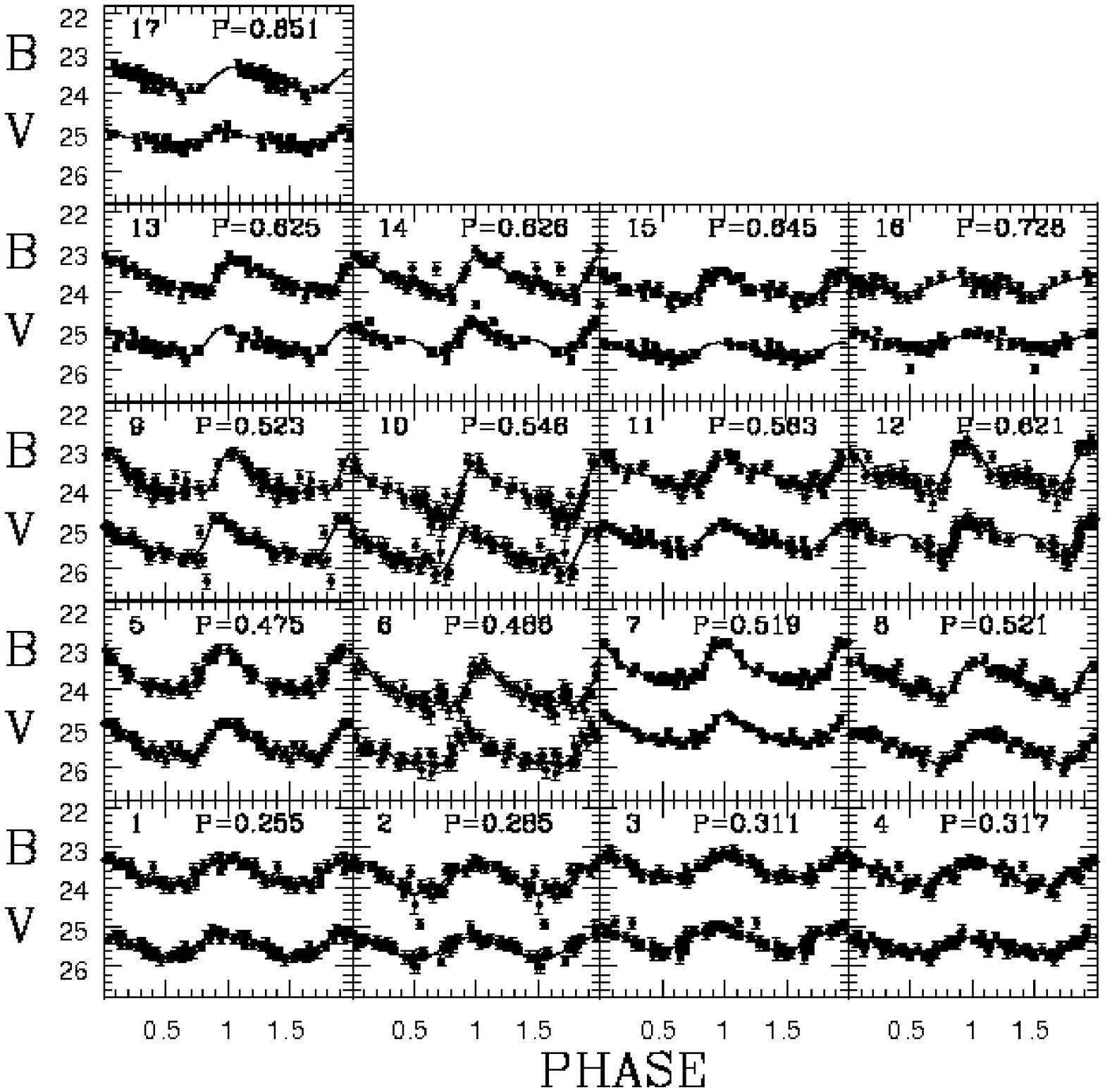}
\caption{Atlas of light curves in the $B$ and $V$ bands of RR Lyrae
  stars detected in F1. The $B$-band points have been shifted brighter
  by 2 mag for clarity.  Error bars, as described in the text, take
  into account both the photometric errors as returned by DOLPHOT
  program as well as the scatter between the data and the model used
  to fit the Fourier series. The model is also shown in this
  Figure. For each variable star, its ID and period obtained by
  fitting the data points are shown in each panel as reported in
  columns 1 and 4 of Table \ref{table:RR_Lyraem32}.}
\label{lc-F1}
\end{figure*}

\begin{figure*}
\plotone{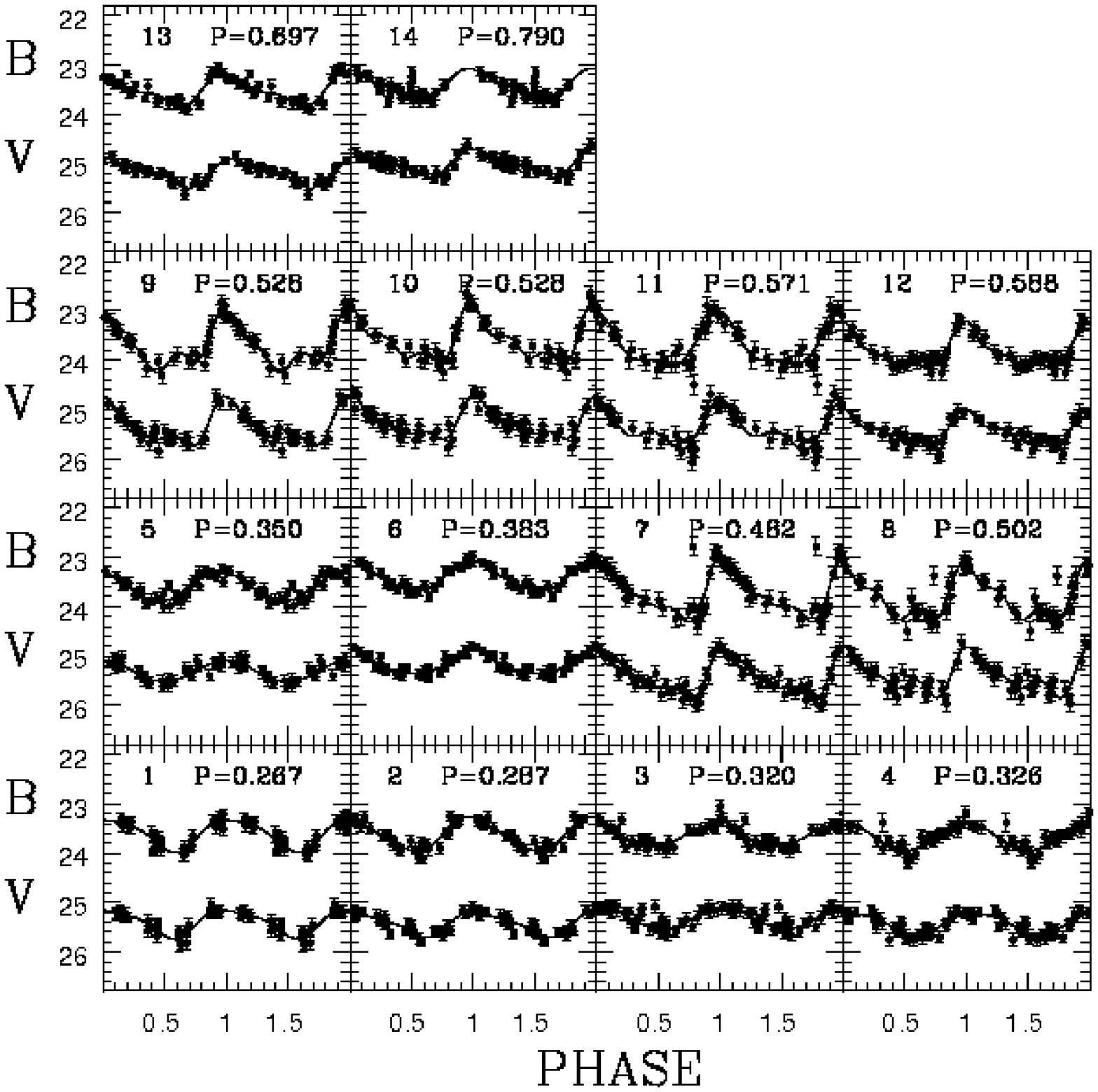}
\caption{As in Figure \ref{lc-F1}, but for RR Lyrae stars detected in
  F2.}
\label{lc-F2}
\end{figure*}

An atlas of the light curves for all the newly-detected RR Lyrae
variables is shown in Figures~\ref{lc-F1} and \ref{lc-F2}. In the
atlas both data points in Johnson-Cousins system as well as the models
used to perform a proper calibration onto this photometric system are
shown. The error bars take into account both the scatter between the
data and the model used to fit the Fourier series (see columns 13 and
14 in Tables~\ref{table:RR_Lyraem32} and \ref{table:RR_Lyraem31}) and
the photometric errors as returned by DOLPHOT program. By averaging
the Johnson-Cousins magnitudes we have obtained $\langle
V\rangle=25.34\pm0.15$ mag for F1 and $\langle V\rangle=25.30\pm0.12$
mag for F2. Then, we have classified RR Lyrae variables into
fundamental-mode (FU) or first-overtone (FO) pulsators by an
inspection of this atlas.  FO pulsators have mean periods of $\sim0.3$
d and sinusoidal light curves, whereas the FU pulsators have longer
periods ( $\langle P_{ab}\rangle=0.59\pm0.11$ d) and more complicated
light curves (with up to 4 harmonics). Because our sample of RR Lyrae
variables is small, we find the same ratios of FO to FU pulsators in
the two fields to within the Poisson errors:
$N_c/N_{\mathrm{total}}=0.23^{+0.27}_{-0.23}$ and
$N_c/N_{\mathrm{total}}=0.42^{+0.58}_{-0.25}$ for F1 and F2,
respectively\footnote{The central estimates and $1\sigma$ confidence
intervals on $N_c/N_{\mathrm{total}}$ have been found using simulations of $10^5$
Poissonian deviates of $N_{\mathrm{total}}$ and $N_c$ in each case.  We use the
median of the resulting distribution of the ratio values as the
central estimate and the region of the diagram that contains 68\% of
the area of the probability distribution function to compute the
$1\sigma$ confidence intervals.}. We discuss the RR Lyrae properties
in detail in Section~\ref{sec:properties}.

\begin{figure}
\plotone{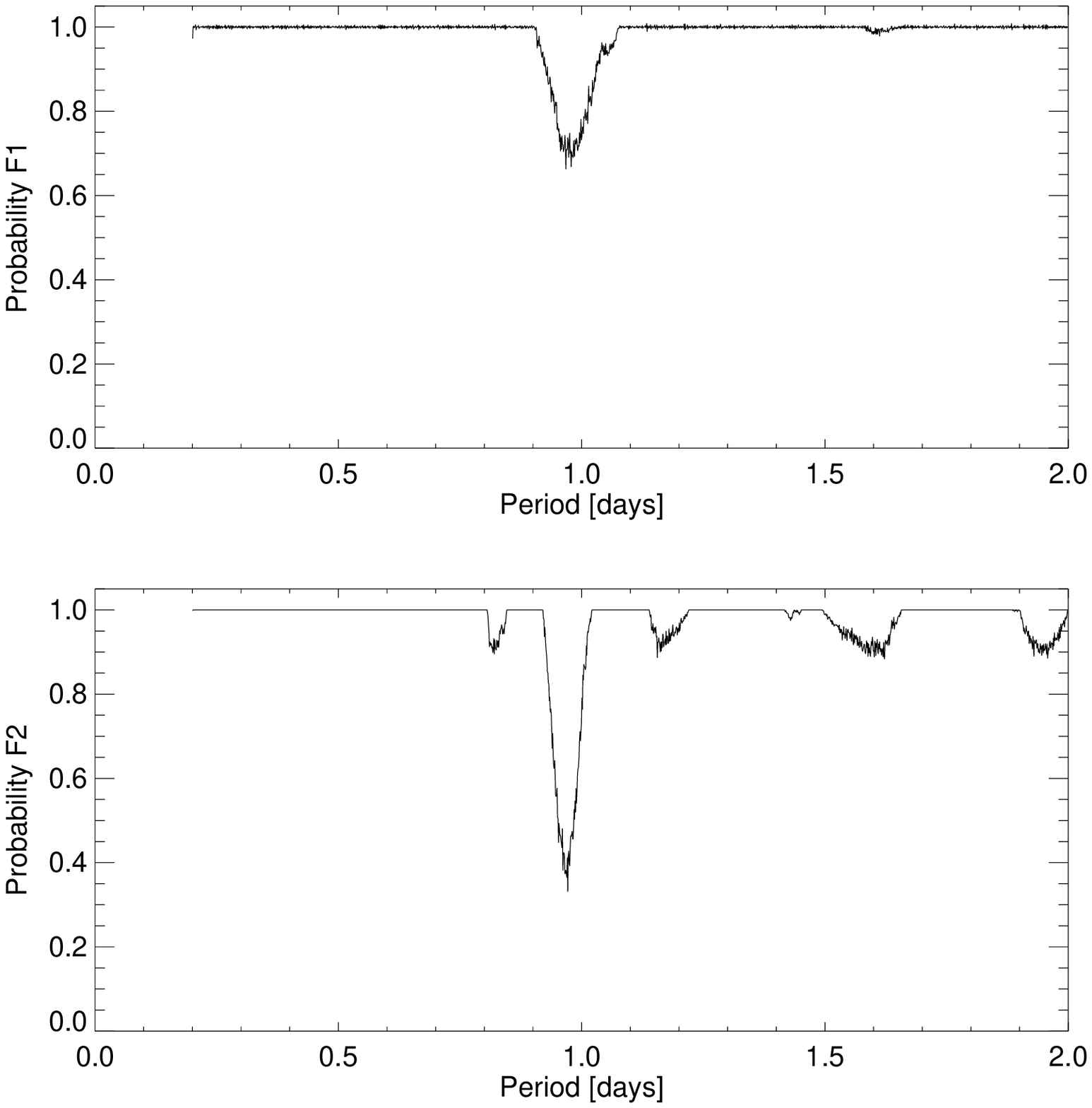}
\caption{Spectra of the probability to detect variability in the
  period range from 0.2--2 d for the F1 (top) and F2 (bottom) fields
  observed with ACS/HRC (this paper). Note that the mean period found
  for F1 and F2 is $0.59\pm0.11$ d.}
\label{p-F1-F2}
\end{figure}

Here we want to conclude by addressing an important question about the
temporal completeness of these observations. That is, could we have
detected all of the RR Lyrae stars in these fields at any reasonable
period? To compute the probability of detecting variability with
periods of 0.2--2 d, we followed the method suggested by
\citet{Saha86} and \citet[see section IV and Figure 7 therein]{SH90},
using software kindly supplied by E. Bernard.  We simulated one
million stars randomly phased and distributed with periods of 0.2--2
days in bins of 0.001 d and then folded the Heliocentric Julian Dates
of both filter datasets according to the random period and initial
phase of each artificial star \citep[see][for details]{Bernard09}.  A
variable is considered recovered if it has a) at least two
observations around the maximum of the light curve, b) at least two
phase points in the descending part of the light curve, and c) a
minimum of three observations during the minimum light. The results
are shown in Figure~\ref{p-F1-F2}, where the probability of detecting
variability is plotted as a function of the period.  The computed
probability for F1 and F2 is nearly unity over the entire range
0.2--0.9 days.  We therefore assume a final (photometric plus
temporal) completeness parameter of 100\% for both fields.

\section{Have We Detected M32 RR Lyrae Variable Stars?}
\label{sec:detection}

We have clearly detected RR Lyrae variables in F1, a field dominated
by M32.  Are any of these stars truly associated with M32?  Or does
the strong background signal from M31 RR Lyrae variables (judging from
F2, which is nearly free of M32 stars) dominate our detection?

We begin addressing these questions by examining the implications of
our detections of RR Lyrae variables in F1 and F2 on the detection of
M32 RR Lyrae variables.  We then extend our analysis to include the
M31 background represented by fields F2--F4 and ask this question
again.  Finally, we examine our results in the context of the study of
\citet{AlonsoGarcia04}, who have previously claimed detection of M32
RR Lyrae variable stars and therefore the presence of an ancient
stellar population in that galaxy.

\begin{deluxetable*}{lllcccccccc}
  \tablecaption{Fields near M32 with claimed RR Lyrae detections} 
  \tablewidth{0pt} 
  \tablecolumns{9}
  \tablehead{ \colhead{ID}& \colhead{RA} & \colhead{DEC} &
    \colhead{Time Window}& \colhead{Instrument}&
    \colhead{N(RR)$_{tot}$} & \colhead{N(RR)$_{FU}$} &
    \colhead{N(RR)$_{FO}$} & \colhead{FoV} & \colhead{Completeness} \\
    & \colhead{(2000)}& \colhead{(2000)} & & &   &  &  &
    \colhead{arcmin$^2$} &  }
    \startdata 
F1\tablenotemark{a}&00:42:47.63&40:50:27.4&2005 Sep 20--24 ($\sim24$ hr)&ACS/HRC&17&13&4&0.25&100\%\\
F2\tablenotemark{a}&00:43:07.89&40:54:14.5&2006 Feb 6--12 ($\sim24$ hr)&ACS/HRC&14&8&6&0.25&100\%\\
F3\tablenotemark{b}&00:42:41.2&40:46:38&2005 Sep 22--24 ($\sim10$ hr)&ACS/WFC&324&267&57&9&97\%\\ 
F4\tablenotemark{b}&00:43:20.8&40:57:25&2006 Feb 9--12 ($\sim10$ hr)&ACS/WFC&357&288&69&9&98\%\\ 
F5\tablenotemark{c}&00:43:01&40:50:21&1998 Nov 19 ($\sim4$ hr)&WFPC2&29&0&0&5.7&5--15\%\\
F6\tablenotemark{c}&00:43:28&41:03:14&1998 Nov 20 ($\sim4$ hr)&WFPC2&16&0&0&5.7&5--15\%
    \enddata 
\tablenotetext{a}{This paper}
\tablenotetext{b}{S09}
\tablenotetext{c}{\citet{AlonsoGarcia04}}
    \label{table:grad}
\end{deluxetable*}

\subsection{M32 RR Lyrae Population Inferred from F1 and F2}

\begin{figure}
  \plotone{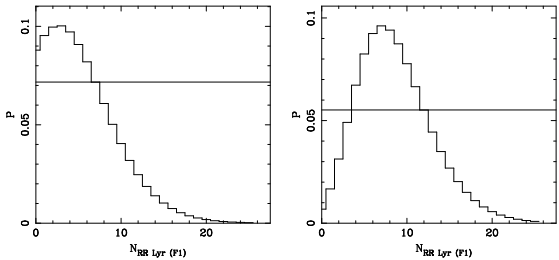}
  \caption{Probability distribution functions for the expectation
    value of the number RR Lyrae variable stars belonging to M32 in
    field F1.  Left:
    $P(N^\mathrm{M32,F1}_{\mathrm{RR\ Lyrae}})=P(\mu_F|N_B+N_F)$
    [equation~(\ref{eq:margF})] using only the number of RR Lyrae
    stars found in F1 and F2.  Right:
    $P(N^\mathrm{M32,F1}_{\mathrm{RR\ Lyrae}})=P(\mu_F|N_B+N_F)$
    [equation~(\ref{eq:margFall})] using the number of RR Lyrae stars
    found in F1--F4, accounting for area differences and completeness.
    See text for more details.\label{probF1RRLyr}}
\end{figure}

We ask the question whether \emph{any} of the RR Lyrae stars in F1
could belong to M32 given the M31 background represented by F2, or at
least what the upper limit on the number of RR Lyrae stars belonging
to M32 is.  We have observed 14 RR Lyrae stars in F2 and 17 in F1.
Our assumption is that the stellar population in F2 represents a
constant background in F1.  Let us call the ``true'' number (the
expectation value) of RR Lyrae stars in each field belonging to M31
$\mu_B$ and the ``true'' number of RR Lyrae stars in F1 belonging to
M32 $\mu_F$.  What we observe is $N_B=14$ in F2 and $N_B+N_F=17$ in
F1.  Then, by Bayes' theorem \citep[see, e.g.][]{Sivia06}, the
probability of finding some $\mu_B$ given $N_B$ is\footnote{Note that
  here we are surpressing the role of the background information $I$,
  so that, for example, $P(N_B)$ is shorthand for $P(N_B|I)$.}
\begin{equation}
  P(\mu_B|N_B) \propto P(N_B|\mu_B) P(\mu_B), \label{eq:probB}
\end{equation}
where the constant of proportionality, $1/P(N_B)$, can be treated as a
normalization constant such that $\int
P(\mu_B|N_B)d\mu_B=1$.  By the product rule, the joint
probability of finding $\mu_F$ and $\mu_B$ given $N_B+N_F$ is
\begin{equation}
  P(\mu_F,\mu_B|N_B+N_F) \propto P(N_B+N_F|\mu_F,\mu_B) P(\mu_B|N_B)
  P(\mu_F), \label{eq:probF}
\end{equation}
where $P(\mu_B)$ and $P(\mu_F)$ are priors on $\mu_B$ and $\mu_F$
which we discuss below, and $P(N_B|\mu_B)$ and
$P(N_B+N_F|\mu_F,\mu_B)$ are both represented by Poisson
distributions, as we are counting stars:
\begin{equation}
  P(N|\mu)=\frac{\mu^Ne^{-\mu}}{N!}.
\end{equation}
In order to determine the probability distribution of $\mu_F$ given
$N_F+N_B$, we need to marginalize equation~(\ref{eq:probF}) over
$\mu_B$:
\begin{equation}
  P(\mu_F|N_B+N_F)\propto\int d\mu_B P(N_B+N_F|\mu_F,\mu_B) P(\mu_B|N_B)
  P(\mu_F). \label{eq:margF}
\end{equation}
All that is left is to specify the priors $P(\mu_B)$ and $P(\mu_F)$
(and to normalize the probability distributions).  Because this is a
location problem \citep[see Chap.~5 of][]{Sivia06}, we choose uniform
priors, with ranges specified by the reasonable range of RR Lyrae
specific frequencies $\mathrm{S_{RR}}$ given the known old populations
in M32 and M31 from M09 [see equation~\ref{eq:SRR} below]:
\begin{equation}
  P(\mu)=\left\{ \begin{array}{ll}
    \frac{1}{b-a} & \textrm{if $a\leq\mu\leq b$}\\
    0 & \textrm{otherwise}
  \end{array}\right., \label{eq:priors}
\end{equation}
where $a$ and $b$ define the range of reasonable values for the
expectation value $\mu$.  For $\mu_F$, we choose $a=0$ and $b=25$, and
for $\mu_B$, we choose $a=1$ to enforce the presence of \emph{some}
background stars in F1 and $b=25$.  These limits on the priors imply
$0\leq\mathrm{S_{RR}}\leq23$ for M32 in F1 and
$1.8\leq\mathrm{S_{RR}}\leq44$ for M31 (assuming the M31 stellar
population is identical in F1 and F2) using the old, metal-poor star
fractions described below.  Note that we require integer numbers of
stars, so the integration over $\mu_B$ in equation~(\ref{eq:margF})
becomes a sum over $\mu_B$, restricted by equation~(\ref{eq:priors})
and our choice of $a$ and $b$ to the range $1\leq\mu_B\leq25$.

With this machinery in place, we find the most probable $\mu_F=3$ RR
Lyrae stars (as expected) belonging to M32, with 68\% confidence
interval of 0--6 RR Lyrae stars (left panel of
Fig.~\ref{probF1RRLyr}).  Another way of looking at this is to assert
that \emph{all} the RR Lyrae variables in F2 come from M31, as stated
above, so $\mu_B=14\pm\sqrt{14}$.  Then the number of \emph{observed}
RR Lyrae variables in F1 is 17, so the number of RR Lyrae variables in
F1 that \emph{do not come from M31} is
$17-\mu_B=17-14\pm\sqrt{14}=3\pm4$ (for integer numbers of stars). We
therefore cannot claim to have detected RR Lyrae stars in M32 with
reasonable confidence based on only F1 and F2, and we can put an upper
limit of no more than 6 RR Lyrae belonging to M32 in F1 from this
analysis.

\subsection{M32 RR Lyrae Population Inferred from F1--F4}

\begin{figure}
  \plotone{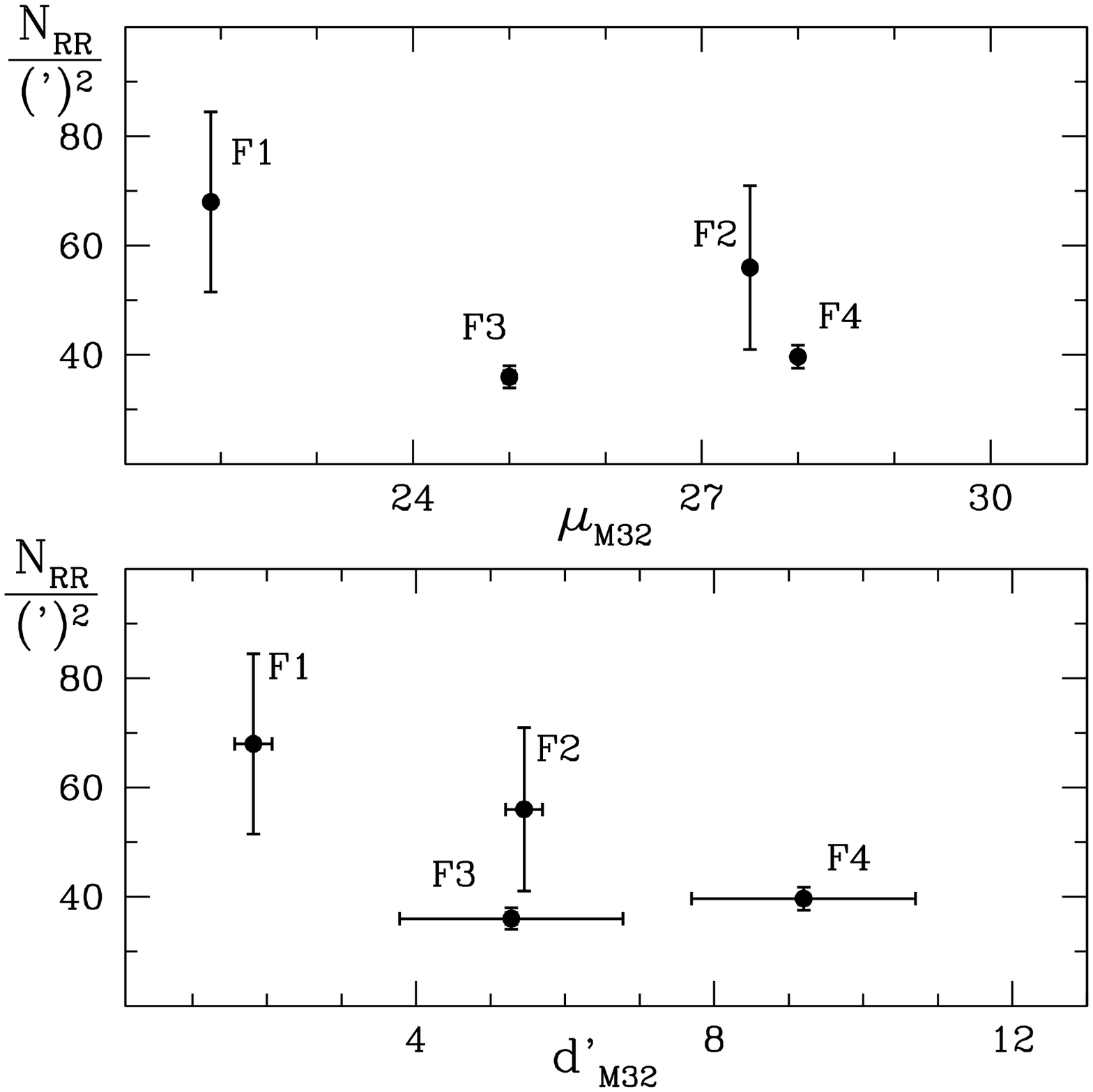}
  \caption{The population of RR Lyrae stars in fields F1--F4 (see
    Fig.~\ref{pointings} for field placements).  Top: RR Lyrae
    population (per square arcminute) as a function of the inferred
    M32 surface brightness (in the $V$ band).  Bottom: RR Lyrae
    population (per square arcminute) as a function of the projected
    distance from M32.\label{NRR}}
\end{figure}

Unfortunately, both fields F1 and F2 suffer from small number
statistics due to the small angular size of the ACS/HRC.  If we allow
that F1 exceeds F2 just by chance, we should also ask if F2 exceeds
the ``true'' background also by chance.  Fields F3 and F4 may provide
some guidance here, as they cover much larger areas (as they were
taken with ACS/WFC) and therefore have much smaller statistical
errors, as shown in Figure~\ref{NRR}. In addition these two fields
nearly fall on the same M31 isophote that passes through F1 and
F2. We therefore continue our search for M32 RR Lyrae variables
assuming that fields F2--F4 constitute a fair sampling of the
background.

\begin{figure}
\plotone{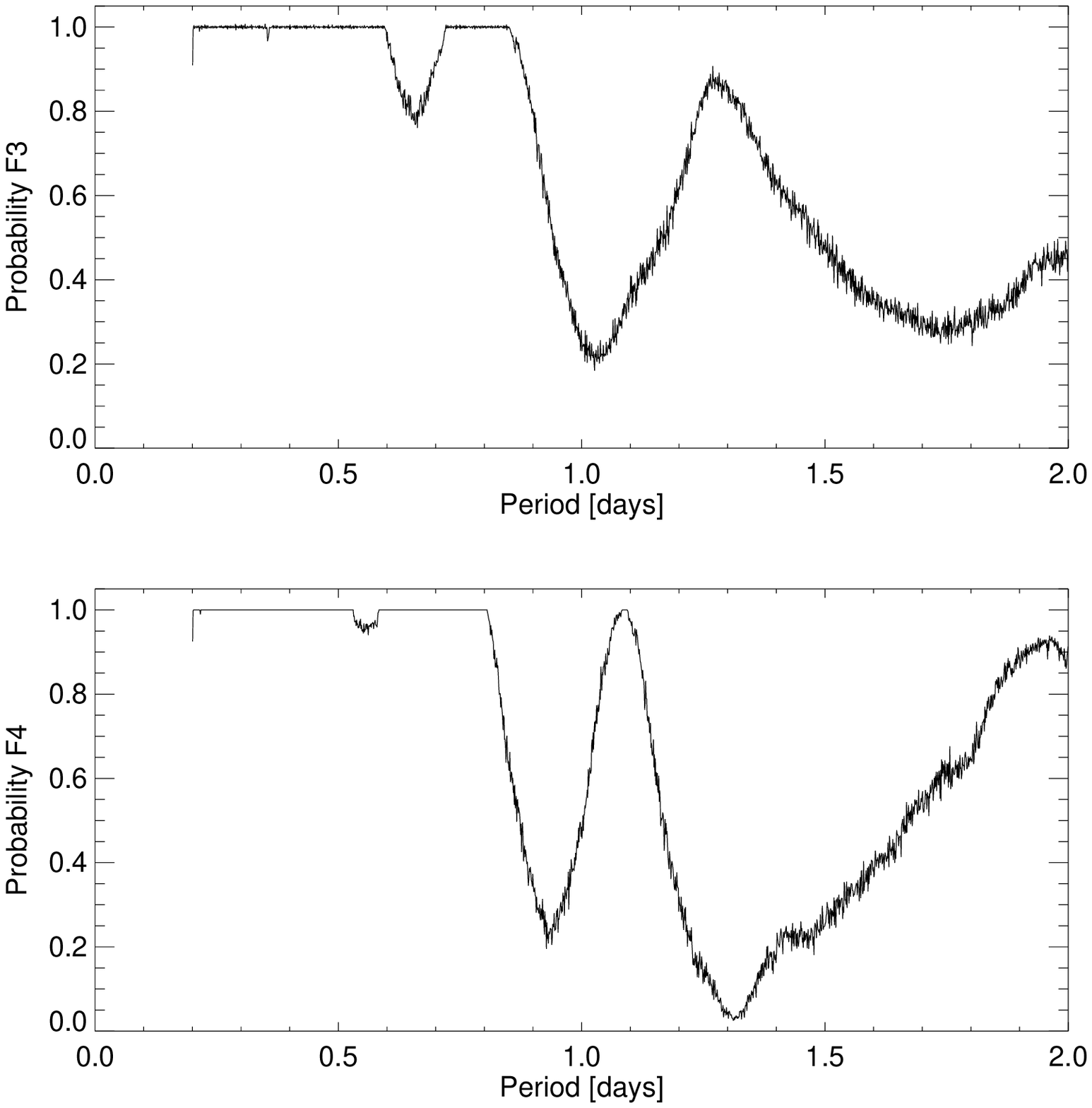}
\caption{Same as in Figure~\ref{p-F1-F2} but for the F3 and F4 fields
  observed with ACS/WFC and analyzed by S09.  Periods found for F3 and
  F4 are $0.55\pm0.07$ d and $0.56\pm0.08$ d, respectively.  These
  spectra are not the same as in Figure~\ref{p-F1-F2} due to the fact
  that these fields were observed for only half of the time that F1
  and F2 were.}
\label{p-F3-F4}
\end{figure}

We must first consider the photometric and temporal completess of F3
and F4. S09 did not perform ASTs on these fields to calculate the
completeness, but according to their luminosity function and to their
comparison with \citet{Brown04}, they assume 100\% photometric
completeness, which we will also assume here.  Using the same
procedure outlined in Section 3, we can estimate the temporal
completeness for these fields.  Figure~\ref{p-F3-F4} shows the
computed probability for F3 and F4, revealing a slight difference
between the sampling of the two fields F3 and F4. In particular for
period values around $\langle P_{ab}\rangle=0.59$ d, the probability
decreases to 0.8 and 0.95 when $P=0.65$ and 0.55 d for F3 and F4,
respectively.  This is due to the shorter time window for these
parallel fields than for the primary fields F1 and F2 (see Sec.~2.1).
There are 45 (out of 324) RR Lyrae variables with periods 0.6--0.7 d
(i.e., 80\% temporal completeness) in F3 and 142 (out of 357) RR Lyrae
variables with periods 0.5--0.6 d (i.e., 95\% temporal completeness)
in F4.  We therefore estimate a total completeness of 97\% and 98\%
for F3 and F4, respectively.

Now we extend our equations~(\ref{eq:probB})--(\ref{eq:priors}) to
include all three background fields, accounting for the different
areas of the background fields and F1.  Because all of the background
field measurements are independent, we can write the number of
background counts as the sum of all three fields, corrected for
completeness:
\begin{equation}
  N_{B^{\prime}}=\sum_{i=\mathrm{F2}}^{\mathrm{F4}}\frac{N_{B_i}}{c_i}=712,
\end{equation}
where $N_{B_i}$ is the number of stars in field $i$ and $c_i$
is the completeness of field $i$.  Then we can write
equation~(\ref{eq:margF}) as
\begin{eqnarray}
  P(\mu_F|N_B+N_F)&\propto&\int d\mu_B P(N_B+N_F|\mu_F,\mu_B)
  \nonumber\\
  & &{}\times
  P\left(\frac{A_{B^{\prime}}}{A_{\mathrm{F1}}}\mu_B|N_{B^{\prime}}\right)P(\mu_F), 
\label{eq:margFall}
\end{eqnarray}
where $A_{B^{\prime}}/A_{\mathrm{F1}}$ is the ratio between the total
area of the background fields and the area of F1,
$A_{B^{\prime}}/A_{\mathrm{F1}} =
\sum_{i=\mathrm{F2}}^{\mathrm{F4}}A_{B_i}/A_{\mathrm{F1}}=73$.

Assuming the same priors as in the previous section
[equation~(\ref{eq:priors}) and the following discussion], we find
that the number of RR Lyrae variables belonging to M32 in F1 has a
most probable value of $\mu_F=7$, with a 68\% confidence interval of
4--11 RR Lyrae variables (right panel of Fig.~\ref{probF1RRLyr}).
Again, we can interpret this result as assuming that the estimated
contribution from M31 RR Lyrae variables in F1 is $712/73=10$, so the
inferred number of M32 RR Lyrae variable in this field is
$17-10\pm\sqrt{10}=7\pm3$ (again, for integer numbers of stars).  Note
that this analysis is only correct if the surface brightness of M31 is
constant across fields F1--F4. We can therefore claim to have detected
seven RR Lyrae variables belonging to M32 in field F1 (with a
$1\sigma$ upper limit of 11 RR Lyrae stars) under this assumption.
Therefore we conclude that, indeed, \emph{M32 has an ancient
population} as represented by the detection at 1-sigma level of RR
Lyrae stars.

\subsection{Specific Frequency of M32 RR Lyrae stars}

We can use this number of RR Lyrae stars belonging to M32 to estimate
an upper limit on the specific frequency of M32 RR Lyrae stars
($\mathrm{S_{RR}}$) in F1, assuming that its old,
metal-poor\footnote{on the basis of the [Fe/H] value as derived in
section \ref{sec:RRFeH}.}  population resembles that of the Galactic
globular clusters. $\mathrm{S_{RR}}$ is defined as the number of RR
Lyrae stars ($\mathrm{N_{RR}}$) normalized to a total Galactic
globular cluster luminosity of $M_{Vt}=-7.5$ mag:
\begin{equation}
\mathrm{S_{RR}}=N_{RR}\,10^{(M_{Vt}+7.5)/2.5}
\label{eq:SRR}
\end{equation}
\citep{Harris96}. We therefore need to obtain the luminosity of the
old, metal-poor population of M32 in our field.  The surface
brightness of M32 in F1 is $\mu_V \approx
21.9\,\mathrm{mag}/\Box\arcsec$ \citep{kfc}. Assuming $E(B-V)=0.08$
\citep{Schlegel98} and a distance modulus $(m-M)_0 = 24.50$ (M09) we
obtain a total luminosity of M32 in F1 of $M_{Vt} \sim -10$ mag.  We
then consider the metallicity distribution function (MDF) of M32
derived by M09, from which a 10 Gyr-old population having $-2.3<
\mathrm{[Fe/H]} < -1.3$ constitutes 11\% of the total luminosity.  The
M32 luminosity of the old, metal-poor population in F1 is thus
$M_{V}=-7.6$ mag. This implies $\mathrm{S_{RR}}\approx6.5$ for this
outer region of M32, with 68\% confidence limits of
$3.6\leq\mathrm{S_{RR}}\leq10$, which is in reasonable agreement with
$\mathrm{S_{RR}}$ of Galactic globular clusters with metallicities
$\mathrm{[Fe/H]} \sim -1.6$ (see Fig.~10 in \citealt{Brown04}).

We can also estimate $\mathrm{S_{RR}}$ for the stellar population of
M31 sampled in F2. The surface brightness in this field is $\mu_V \sim
22.7\,\mathrm{mag}/\Box\arcsec$, and assuming the same distance
modulus and reddening than F1, we obtain a total luminosity of M31 in
F2 of $M_{Vt} \sim -9.24$ mag.  We consider again only the old,
metal-poor population in F2 and calculate its luminosity. From the MDF
of M31 by M09, a 10 Gyr-old population with metallicities $-2.3<
\mathrm{[Fe/H]} < -1.3$ constitutes a 12\% of the total luminosity,
which translates into $M_{Vt} \sim -6.9$ mag. The 14 RR Lyrae stars
found in F2 imply a $\mathrm{S_{RR}} \sim 18$ of M31 in this field.
Note that this value is higher than $\mathrm{S_{RR}} \sim 11.2$
estimated by \citet{Brown04} for their M31 halo field. However, the
scatter in $\mathrm{S_{RR}}$ in the Galactic globular cluster system
is large enough to encompass these variations if M31's RR Lyrae
population is similar to the Milky Way's \citep[cf.\ the discussion
  in][]{Brown04}.

One might be tempted to attempt to invert this analysis to determine
the (upper limit on the) fraction of old, metal-poor stars in M32 and
M31.  After all, we know the upper limit on the number of RR Lyrae
stars we can associate with M32 in F1 and the number we can associate
with M31 in F2.  However, no theoretical or empirical one-to-one
relationship between $\mathrm{S_{RR}}$ and metallicity exists, due to
the scatter in this $\mathrm{S_{RR}}$--[Fe/H] plane induced by the
``second-parameter problem'' \citep[see, e.g.,][]{Suntzeff91,Brown04}.
Furthermore, we have used the \emph{known} fraction of metal-poor
stars in M32 in F1 (and in M31 in F2) determined by M09 and a
reasonable guess for the ages of these metal-poor stars to estimate
$\mathrm{S_{RR}}$, so such an analysis would be circular.
Alternately, if there were a stellar evolution model that correctly
predicted mass-loss along the first-ascent giant branch so that the
distribution of stars along the horizontal branch was also correctly
predicted, we could also perform this inversion; but no such model
currently exists \citep[see, e.g.,][]{SCbook}.

\subsection{Comparison with Alonso-Garc{\'\i}a et al.~(2004)}

\citet{AlonsoGarcia04} were the first to attempt to directly detect an
ancient population in M32 using RR Lyrae variables. They claimed to
have detected $12\pm8$ RR Lyrae stars in a field further away from M32
than F1.  From this claimed detection, they conclude that 2.3\% of
M32's stellar population is old.

\begin{figure}
\plotone{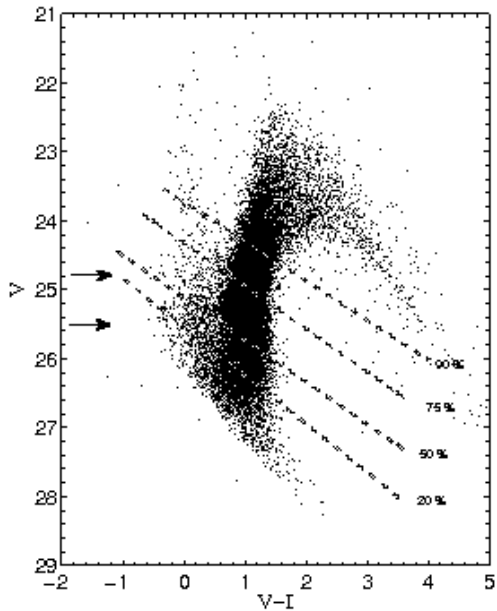}
\caption{The JC-calibrated ($V-I$,$V$) CMD for field F5 and
  various completeness levels (dashed lines). The arrows show the
  location where we expect to find RR Lyrae stars. On this basis we
  conclude that the photometric completeness is between 20 and
  75\%. See M09 for more details.}
\label{comp}
\end{figure}

\begin{figure}
\plotone{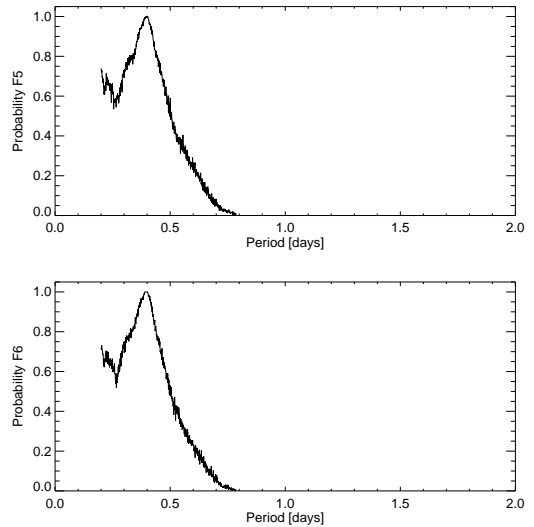}
\caption{Same as in Figure~\ref{p-F1-F2} but for the F5 and F6 fields
  observed with WFPC2 and analyzed by \citet{AlonsoGarcia04}.}
\label{p-F5-F6}
\end{figure}

In this section, we analyze the temporal and photometric completeness
of Alonso-Garc{\'\i}a et al.'s data in order to determine if our
detections and theirs are compatible. The authors do not give any
estimate of the photometric completeness, but by examining their CMD
we can see that the luminosity level of the horizontal branch is
evidently at the detection limit of the WFPC2 exposures. We call their
primary (``M32'') field F5 and their background (``M31'') field F6 in
Table~\ref{table:grad}.  M09 performed photometry and ASTs on F5 to
compute their completeness levels using HSTphot \citep{Dolphin00}. In
Figure~\ref{comp} we show the ($V-I$, $V$) CMD calibrated onto the JC
photometric system using the calibrations provided by
\citet{Holtzman95}. We conclude that the photometric completeness for
RR Lyrae stars in F5 is 20--75\%, depending strongly on color.  Field
F6 is similarly deep and therefore we assume that its photometric
completeness is identical.  Figure~\ref{p-F5-F6} shows that the
temporal completeness computed for F5 and F6 using the method
described in Section~3 is only $\approx0.2$ for the mean period
observed for RRab Lyrae ($P\sim0.59$ days) and is 0.6--0.8 for RRc
($P\sim0.28$--0.32 days). We stress here that the RRc detectability
mostly depends on the quality of photometry due to the very low
amplitudes of FO pulsators and that \citet{AlonsoGarcia04} likely
could not have detected any of these. Thus we can assume a total
(photometric and temporal) completeness for these fields of around
5--15\%.

Given our own RR Lyrae detections in F1, we can estimate their
completeness in a different manner.  We note that their WFPC2 fields
are 22.8 times larger in area than our ACS/HRC fields, but that the
M32 surface brightness in F5 is
$\mu^{\mathrm{M32,F5}}_V\sim23.7\,\mathrm{mag}/\Box\arcsec$, about 5.2
times fainter than F1.  We assumed that the M31 surface brightness is
the same in both fields,
$\mu^{\mathrm{M31}}_V\sim22.7\,\mathrm{mag}/\Box\arcsec$.  Then the
total number of RR Lyrae we predict that \citet{AlonsoGarcia04} should
have detected if they had 100\% completeness, given
$N_{\mathrm{M31,F1}}$ RR Lyrae variables belonging to M31 and
$N_{\mathrm{M32,F1}}$ RR Lyrae variables belonging to M32 in F1, is
\begin{equation}
  N_{\mathrm{tot,F5}}=22.8 N_{\mathrm{M31,F1}}+22.8
  N_{\mathrm{M32,F1}} \frac{10^{-0.4\mu_V(\mathrm{M32,F5})}}{10^{-0.4\mu_V(\mathrm{M32,F1)}}}.
\end{equation}
For $N_{\mathrm{M32,F1}}=7_{-3}^{+4}$ RR Lyrae variables belonging to
M32 in F1, they should have detected
$N_{\mathrm{tot,F5}}=258_{+55}^{-74}$ RR Lyrae in total (with fewer
detected in F5 for more M32 variables in F1).  They claim to have
detected 22 RR Lyrae variables in F5, so we estimate their
completeness to be $c_{\mathrm{F5}}=0.085_{-0.015}^{+0.034}$.
Due to this severe incompleteness, \citet{AlonsoGarcia04} were not
able to study the intrinsic properties of the stars they detected. In
fact, they classified variable stars detected as RR Lyrae stars due
only to their location in the CMD.

We can now compute the number of M32 stars we should have seen in F1
if \citet{AlonsoGarcia04} detected $12\pm8$ RR Lyrae variables
belonging to M32 in F5:
\begin{equation}
  N_{\mathrm{RR,M32}}(\mathrm{F1}) = N_{\mathrm{RR,M32}}(\mathrm{F5}) \frac{A_{\mathrm{F1}}}{A_{\mathrm{F5}}}\frac{10^{-0.4\mu_V(\mathrm{F1})}}{10^{-0.4\mu_V(\mathrm{F5)}}}\frac{c_{\mathrm{F1}}}{c_{\mathrm{F5}}},
\end{equation}
where $A_{\mathrm{F1}}$ and $A_{\mathrm{F5}}$ are the areas of the two
fields, $\mu_V(\mathrm{F1})$ and $\mu_V(\mathrm{F5})$ are their
surface brightnesses, and $c_{\mathrm{F1}}$ and $c_{\mathrm{F5}}$ are
the completeness levels at the blue horizontal branch in the two
fields.  We assume here that this last ratio is $1/0.085$, based on
the completeness estimate above.  The scaling factor in this equation
is then 2.7, a number which is driven mostly by the incompleteness of
F5.  That is, we should have seen 2.7 times as many M32 RR Lyrae
variables in F1 than \citet{AlonsoGarcia04} saw in F5, for a total of
$33\pm22$ RR Lyrae variables predicted to belong to M32 in our field.
This is a factor of 4.6 higher than the number that we claim to have
detected.  However, the lower 68\% confidence level of this estimate
just overlaps with the upper 68\% confidence level of our estimate of
$7_{-3}^{+4}$ RR Lyrae variables predicted to belong to M32 in F1.  We
can of course invert this procedure and predict how many M32 RR Lyrae
variables \citet{AlonsoGarcia04} should have seen in F5 if our
detection is correct.  We predict that they should have seen
$3_{-2}^{+3}$ RR Lyrae variables predicted to belong to M32 in F5,
once again just in agreement (within the 68\% confidence intervals)
with their estimate of $12\pm8$.  We therefore suggest that it is
possible that \citet{AlonsoGarcia04} detected \emph{bona fide} M32 RR
Lyrae variables in F5; but their severe incompleteness makes their
detection highly uncertain.

\smallskip
\smallskip
We summarize this section by stating that considering only our ACS/HRC
fields F1 and F2, we have detected $\leq6$ RR Lyrae variable stars
belonging to M32 in F1. Complementing our primary observations with
our ACS/WFC parallel fields F3 and F4 allows a better determination of
the likely M31 background in F1, and using this estimate, we claim to
have detected $7_{-3}^{+4}$ RR Lyrae variable stars belonging to M32
in F1, with a most probable value of seven variable stars.  We
therefore conclude that M32 has an ancient population that can be
detected without directly probing the oldest main-sequence turnoffs.

\section{RR Lyrae Properties}
\label{sec:properties}

We now ask whether we can separate the detected RR Lyrae stars
belonging to the M32 and M31 stellar populations based on their
intrinsic pulsation properties.

\subsection{RR Lyrae Star Colors and the Pulsation Instability Strip}
\label{sec:PIS}

\begin{figure}
\plotone{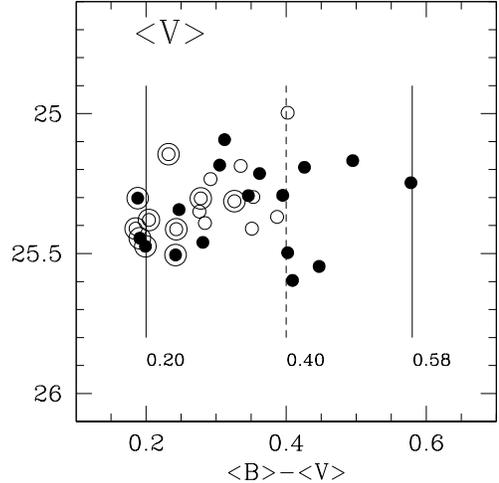}
\caption{Zoomed view of the CMD around the positions of RR Lyrae
  stars.  Mean magnitudes have been calibrated onto the Johnson-Cousin
  photometric system. RR Lyrae stars from field F1 and those from F2
  are displayed as filled and empty circles, respectively. FO
  pulsators are emphasized by using other larger circles around the
  symbols used. The boundaries of the pulsation Instability Strip and
  the relative values we adopted to draw them are indicated in each
  panel, namely $\mathrm{FOBE}_{\mathrm{F1}-\mathrm{F2}}=0.20$ mag
  (left solid line), $\mathrm{FRE}_{\mathrm{F2}}=0.40$ mag (dashed
  line) and $\mathrm{FRE}_{\mathrm{F1}}=0.58$ mag (right solid
  line). Interestingly RR Lyrae stars in F1 appear redder than the
  ones in F2 suggesting a large spread in metallicity in this field
  (see the text for details).}
\label{strip}
\end{figure}

Does the average color of the RR Lyrae stars allow us to separate the
two populations?  In Figure~\ref{strip} we show a blow-up of the CMDs
in $V$ calibrated onto the JC photometric system.  Inspecting this
figure we see a clear difference in the color $(\langle
B\rangle-\langle V\rangle)$ range of variables belonging to our
different fields. In particular, F1 RR Lyrae colors are spread through
a larger color range (0.18 mag) than those in F2.

To understand this difference in color range, we need to discuss the
mechanisms driving the radial stellar pulsation. There are two main
pulsation mechanisms, both related to the opacity in the stellar
envelope layers where partially-ionized elements are abundant
(hydrogen and neutral or singly-ionized helium): the $\kappa$- and
$\gamma$- mechanisms\footnote{These mechanisms are triggered in a
  region where an abundant element (hydrogen or helium) is partially
  ionized \citep[see][]{Zhevakin53,Zhevakin59,BK62,Cox63,BS94}.
  During an adiabatic compression the opacity increases with the
  temperature \citep{KC68}, which is unlikely if the element is
  completely ionized.  During this compression, heat is retained and
  the layer contributes to the instability of the structure. This is
  called the $\kappa$-mechanism and was first discussed by
  \citet{Eddington26}, who called it the ``valve mechanism.''  The
  $\gamma$-mechanism allows, under an initial compression, the energy
  to be absorbed by the ionizing matter, instead of raising the local
  temperature, therefore lowering the adiabatic exponent $\gamma$. The
  layer then tends to further absorb heat during compression, leading
  to a driving force for the pulsations \citep{KC68}. It is evident
  that the two mechanisms are connected with each other. H, HeI and
  HeII layers are located at 13,000 K, 17,000 K, and 30,000--60,000 K,
  respectively \citep[and references therein]{BS94}, and thus this
  phenomenon involves only the stellar envelope.}. These are directly
correlated to the variations, in these layers, of the opacity
($\kappa$) and the adiabatic exponent
($\Gamma_3-1$\footnote{$\Gamma_3-1=\delta\log T/\delta\log\rho$}),
respectively.  These mechanisms can explain the pulsational
instability strip (IS), the well-defined region in color where we find
RR Lyrae stars and radial periodic pulsators in general (as classical
and anomalous Cepheids, SX Phe variables, etc.) are observed. If a
low-mass star ($M\la 0.8 M_{\odot}$) in the core helium- and
hydrogen-shell-burning phase (the zero-age horizontal branch) crosses
the IS during its evolution, it becomes a RR Lyrae variable star.

Stars bluer than the First Overtone Blue Edge (FOBE) of the IS cannot
pulsate because the ionization regions (located at almost constant
temperatures) are too close to the surface, and thus the stellar
envelope mass is too low to effectively retain heat and act as a
valve. Moving towards the red side, at lower effective temperatures,
convection becomes more efficient and, at the Fundamental Red Edge
(FRE), quenches the pulsation mechanism \citep{Smith95}. The width of
the IS is then just the color difference between the FOBE and the FRE.
The FOBE location in the CMDs is well defined by theory, and we use
this in Section 5.3.2 to find an independent estimate of the distance
modulus \citep[as suggested by][]{Caputo97,Caputo00}, while the
position of the FRE depends on the assumptions adopted to treat the
convective transport, becoming bluer as the convective efficiency is
increased.

The IS depends on the initial stellar chemical composition due to the
dependence of the pulsation physics on the opacity and on the hydrogen
and helium partial-ionization regions. In particular the metallicity
and the convection efficiency (especially for the FRE) should have the
largest effects on the boundaries of the IS. As the initial stellar
metallicity increases, the opacity increases, resulting in a more
efficient $\kappa$-mechanism at lower effective temperature, which
implies a redder FRE. The FOBE location is almost constant, as it
depends essentially on the temperature of the region where helium is
completely ionized \citep[see, e.g.,][]{Walker98,Caputo00,Fiorentino02}.

In our case, shown in Figure~\ref{strip}, we do not see any difference
in the FOBE between the fields, with $(\langle B\rangle-\langle
V\rangle)_{\mathrm{FOBE}}=0.2$ mag for both fields, whereas $(\langle
B\rangle-\langle V\rangle)_{\mathrm{FRE}}=0.58$ mag for F1 and $0.40$
mag for F2. This evidence could be interpreted as a suggestion that F1
RR Lyrae field stars may have a larger spread in metallicity or age
than their M31 counterparts, support for two different mixes of
stellar populations in field F1. Furthermore it is interestingly to
note these red RR Lyrae stars (namely V6, V10, V14, V15, V16 and V17)
in F1 seem to be closer to the M32 center, located as indicated by the
arrow in Figure~\ref{FC-F1}. Note also that the F2 RR Lyrae stars V3
(the reddest RR$_c$) and V7 (the bluest RR$_{ab}$) define an
empirical ``OR'' region of the pulsation IS (see
Figures~\ref{cmd-F1F2zoom} and in ~\ref{strip}) where, depending on
their evolutionary path, RR Lyrae stars may pulsate in both FU or FO
mode.

\subsubsection{RR Lyrae Reddening Evaluation}
\label{sec:RE}

To understand whether this difference in the color spreads is
intrinsic---that is, due to the metallicities or ages of the
stars---or just a reddening effect, we must study the intrinsic
reddening. From \citet{Schlegel98} we know that Galactic
\emph{foreground} extinction in that direction is $E(B-V)=0.08\pm0.03$
mag and is essentially the same for both fields F1 and F2.  However,
we must also allow for extinction from the disk of M31, in case some
of the RR Lyrae variables in F1 and F2 lie behind it. In the following
we will use the intrinsic properties of the RR Lyrae stars to derive
the \emph{total} intrinsic reddening values for both fields in two
independent ways.

The first method is based on the insensitivity of $\langle (B - V)
\rangle _{\mathrm{FOBE}}$ to the intrinsic properties of different RR
Lyrae populations, stressed for the first time by \citet{Walker98}.
Walker presents dereddened (magnitude-averaged) $\langle (B-V)\rangle$
IS boundary colors from accurate observations of nine Galactic and LMC
globular clusters covering a range in metallicity from
$\mathrm{[Fe/H]}= -1.1$ to $-2.2$ dex.  He concludes that the
dereddened color of the blue edge is at $\langle (B-V)\rangle
_{\mathrm{FOBE}}=0.18\pm0.02$ mag, with no discernible dependence on
metallicity, while the color of the red edge shows a shift of
$0.04\pm0.03$ mag.  This method was also used successfully by
\citet{Clementini03} to estimate the average reddening value in LMC RR
Lyrae stars. As discussed above both fields have the same FOBE color,
and therefore we obtain the same estimate for their reddening.  We
find a reddened color of $(\langle B\rangle-\langle
V\rangle)_{\mathrm{FOBE}}$\footnote{In our sample (F1 and F2) we
verified that no significant difference is found between the FOBE as
derived by the magnitude and intensity averaged colors, i.e. $(\langle
B\rangle - \langle V\rangle)_{\mathrm{FOBE}} - \langle (B-V)\rangle
_{\mathrm{FOBE}} = 0.00 \pm 0.02$ mag.}$=0.20\pm0.03$ mag for both
fields, as shown in Figure~\ref{strip}. Thus, we infer a value of
$E(B-V)=0.02\pm0.04$ mag for both fields.  Interestingly, this value
of the \emph{total} reddening is more than $1\sigma$ \emph{smaller}
than the estimate from the \citet{Schlegel98} map, which is meant to
measure the \emph{foreground} reddening.

The second method was described originally by \citet{Sturch66}. We
stress here the importance of this method, as the calibration of the
\textsc{H i} column densities with reddening is established on its
basis and used by \citet{BH78,BH82}. These authors assumed an offset
of $E(B-V)=0.03$ from the values found by \citet{Sturch66}. They fixed
it under the assumption that at the Galactic poles $E(B-V)=0$, as
suggested by \citet{McDonald77} in his analysis of RR Lyrae colors and
H$\beta$ indices, whereas Sturch found $E(B-V)=0.03$.  With this
method we derive reddening for each RRab star from the
(magnitude-averaged) color at minimum light $(B-V)_{min}$ (phases
between 0.5 and 0.7), the period $P_{ab}$, and the metal abundance
[Fe/H] of the variables. The application of Sturch's method requires
the knowledge of the metallicity of each individual RRab. Because we
do not have this information from spectroscopy of these fields, we
have decided to fix the average metallicity for both field to the
value $\mathrm{[Fe/H]}= -1.6$ dex, as discussed in
Section~\ref{sec:RRFeH}. Sturch's method was calibrated on Galactic
field RR Lyrae stars \citep{Sturch66,Blanco92} and has been used on
Galactic Globular clusters \citep{Walker90,Walker98} and on LMC RR
Lyrae stars \citet{Clementini03}, returning, as expected, very good
agreement with the \textsc{H i}-based reddening measurements taking
into account the \citet{BH78,BH82} extinction zero point.
\citet{Walker92} used the following formulation of the Sturch's
method,
\begin{equation}
  E(B-V)=(B-V)_{min}-0.336-0.24P(\mathrm{d})-0.056\mathrm{[Fe/H]},
\end{equation}
where the reddening zero point has been adjusted to give $E(B-V) =
0.0$ mag at the Galactic poles, and the [Fe/H] is that of the Zinn \&
West (1984) metallicity scale. We infer mean reddening values of
$E(B-V)=0.07\pm0.10$ mag and $0.03\pm0.08$ mag for F1 and F2
respectively, in statistical agreement with each other and with the
\citet{Schlegel98} map.  Note however that the scatter in the actual
metallicities of the RR Lyraes will increase the scatter in these
estimates.

Depending on the assumed method, we therefore find two different
estimates of the reddening value: a) the \citet{Schlegel98} map and
the FOBE color method suggest the same reddening for both fields but
different values depending on the method, 0.08 mag and 0.02 mag
respectively; b) Sturch's method suggests a reddening slightly higher
for F1 field than for F2 of about 0.04 mag. Taking into account the
errors on both methods we do not find any significant difference
between the two evaluations as well as between the two fields.

We will discuss all these cases when determining the distance modulus:
the first one assuming $E(B-V) = 0.08$ for both fields by following
the \citet{Schlegel98} map, the second one assuming $E(B-V)=0.02$ mag
for both fields as suggested by the FOBE method, and the third one
assuming different values, e.g., $E(B-V)=0.07$ mag for F1 and
$E(B-V)=0.03$ mag for F2 found using Sturch's method. In all cases an
extinction law with $R_V=3.1$ will be assumed.

\subsection{Mean Periods, Amplitudes and Oosterhoff Types}

In this section we focus our attention on their mean periods,
amplitudes and Oosterhoff types (Oo types) in order to distinguish
between the two populations, M32 and M31 halo (and/or outer disk)
variables. Pulsational periods and amplitudes are of fundamental
importance because they depend on the star structural parameters
(mass, luminosity and effective temperature) and are distance- and
reddening-free observables.

RR Lyrae variables found in Galactic globular clusters can be divided
in two distinct classes by the mean periods of their fundamental mode
pulsators: Oosterhoff type I and II \citep{Oosterhoff39}.  In
Oosterhoff type I (OoI) Galactic globular clusters, RRab variables
have average periods of $\langle P_{ab}\rangle = 0.559$ d, and in
Oosterhoff type II (OoII) Galactic globular clusters they have average
periods of $\langle P_{ab}\rangle = 0.659$ d \citep{Clement01}, very
few clusters have RRab with periods in the range of $0.58 \leq \langle
P_{ab}\rangle\leq 0.62$ d (the ``Oosterhoff gap''). These different
Galactic Oosterhoff types may have resulted from different accretion
and formation processes in the Galactic halo, a hypothesis supported
by the difference in metallicities found for the two types (OoII RR
Lyrae are on average more metal poor than OoI RR Lyrae, with
$\langle[\mathrm{Fe/H}]\rangle\approx-1.8$ and $\approx-1.2$,
respectively, although with large and overlapping metallicity
distributions: \citealt{SPP09}).  The Oosterhoff dichotomy may well
hold a key to the formation history of the Galactic halo, but much
evidence has suggested that this dichotomy cannot support the galaxy
formation scenario in which present-day dwarf spheroidal (dSph)
satellite galaxies of the Milky Way are the building blocks of our
Galaxy \citep[see, e.g.,][]{Catelan06}. In fact the RR Lyrae variables
of these dSphs as well as those of their globular clusters fall
preferentially into the ``Oosterhoff gap.''  Furthermore,
\citet{Clement01} investigated the properties of the RR Lyrae
variables in Galactic globular clusters and found that the ratio
between the number of RRc and RRab stars is constant for each of the
two Oosterhoff types: $N_{c}/N_{\mathrm{total}}=0.22$ for the OoI clusters and
$N_{c}/N_{\mathrm{total}}=0.48$ for the OoII clusters.  It is currently not clear
whether the Oo dichotomy is a peculiarity of our own Galaxy; for
example, the RR Lyrae stars observed in the Magellanic Clouds do not
appear to follow this dichotomy \citep[see][and references
therein]{Alcock00}.

  \citet{Brown04} claimed that the RR Lyrae population they discovered
  in the M31 halo cannot be classified into either of the Oo types.
  They found a mean period of $\langle P_{ab}\rangle = 0.60\pm0.10$ d
  (in the Oo gap) but $N_{c}/N_{\mathrm{total}}=0.46$, higher than the
  typical value for OoI clusters (0.22). On the other hand,
  \citet{Clementini09} found that one of the most luminous M31
  clusters, B514, shows a peculiarity in the Oosterhoff
  dichotomy. They found that the mean period of 82 RRab stars is
  $\langle P_{ab}\rangle=0.58$ d and $N_{c}/N_{\mathrm{total}}=0.08$,
  suggesting an OoI type, while at the same time having a very low
  metallicity, $\mathrm{[Fe/H]}\sim-1.8$ \citep{Galleti06}, suggesting
  an OoII type. The very low value of the $N_{c}/N_{\mathrm{total}}$
  ratio in this cluster found by \citet{Clementini09} can be explained
  by the incompleteness of their sample due to the intrinsically low
  amplitudes of RRc stars.  In a recent study based on parallel
  observations of our program taken with ACS/WFC, S09 found $\langle
  P_{ab}\rangle=0.55 \pm 0.07$ d and $N_c/N_{\mathrm{total}} = 0.24$
  and $\langle P_{ab}\rangle=0.56 \pm 0.08$ d and
  $N_c/N_{\mathrm{total}} = 0.21$ for the F3 and F4 fields (see
  Table~\ref{table:grad}), respectively, suggesting an OoI type
  classification.  S09 applied the period--metallicity--amplitude
  relation to their large sample (almost 700 stars) and found
  $\mathrm{[Fe/H]}=-1.77\pm0.06$ for both fields, similar to the
  \citet{Clementini09} results.

As briefly discussed in Section \ref{sec:identification}, we find the
same mean period $\langle P_{ab}\rangle=0.59\pm0.11$ d and, within the
uncertainties, the same ratios of FO to FU pulsators in the two
fields: $N_c/N_{\mathrm{total}}=0.23^{+0.27}_{-0.23}$ and
$N_c/N_{\mathrm{total}}=0.42^{+0.58}_{-0.25}$ for F1 and F2,
respectively. The error on the mean periods is just the standard
deviations of the average value computed on our very small sample of
RRab Lyrae stars, i.e., 13 stars for F1 and 8 for F2, and the errors
on the ratio of FO to FU pulsators reflect only Poissonian counting
uncertainties.

\begin{figure}
\plotone{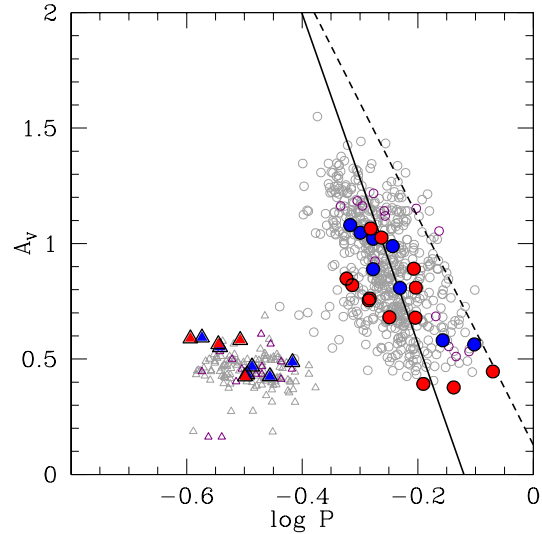}
\caption{Bailey Diagram ($V$-band amplitude as a function of period)
  for RR Lyrae stars near M32 and in the disk of M31. RR Lyrae stars
  in F1 are shown in red, those in F2 are shown in blue, those found
  by S09 in F3 and F4 are in grey, and those found in \citet{Brown04}
  in F7 are in magenta.  FU pulsators are circles and FO pulsators are
  triangles.  As expected the FO pulsators are concentrated in a
  region with low amplitude and periods, whereas the FU pulsators show
  a linear behavior, i.e., their amplitudes decrease as their periods
  increase. That trend is stressed by the solid and dashed lines
  representing Oosterhoff types I and II, respectively
  \citep{Clement00}. Nearly all FU pulsators follow the OoI relation.}
\label{bailey}
\end{figure}

We ask whether these properties are consistent with a single
Oosterhoff type. In Figure~\ref{bailey} we show the composite Bailey
diagram of \citet{Brown04}, S09, and our new detection of RR Lyrae
variables. The comparison with other samples is made by assuming that
their amplitudes, measured in the broad $F606W$ filter (with a peak
wavelength between Johnson-Cousins the $V$ and $R$ bands),
underestimate the JC $V$-band amplitudes by $\approx8$\% \citep[for
  further details see][]{Brown04}.  Inspecting this figure, we see
that RR Lyrae stars in both of our fields can be classified as OoI, as
expected given the metallicity distribution of these fields (M09), and
in particular, their lack of well-developed blue horizontal branches.

\subsection{RR Lyrae Pulsation Relations}

We next attempt to disentangle the putative M32 RR Lyrae population
from that of the M31 using various observational manifestations of the
theoretically-predicted RR Lyrae period--luminosity--temperature
relation valid for every \emph{individual} RR Lyrae star.

\subsubsection{RR Lyrae Metallicities: [Fe/H]--Period--Amplitude
  Relation}
\label{sec:RRFeH}

To estimate the metallicity of the RR Lyrae stars in the Large
Magellanic Cloud (LMC), \citet{Alcock00} determined a relation between
the metallicity [Fe/H] of an RRab variable, its period $P_{ab}$, and
its $V$-band amplitude $A_V$,
\begin{equation}
\mathrm{[Fe/H]}_{ZW} = -8.85[\log P_{ab} + 0.15 A_V]-2.60,
\label{eq:FeH}
\end{equation}
where $ZW$ refers to the \citet{ZW84} metallicity scale.  This
relation has been calibrated from Galactic globular clusters with
metallicities in the range of $-1.4\leq\mathrm{[Fe/H]}_{ZW}\leq-2.1$
dex (and has been checked against the metallicities of Galactic field
RR Lyraes).  The uncertainty of this relation is
$\sigma_{\mathrm{[Fe/H]}}=0.31$.  This relation shows very good
agreement with independent metallicity estimations for the LMC
\citep{Alcock00}, And~IV \citep{Pritzl02}, and And~II
\citep{Pritzl04}.  Applying this relation to the RR Lyrae variables in
our two fields we find mean values of
$\langle\mathrm{[Fe/H]_{F1}}\rangle = -1.52 \pm 0.10 \pm 0.48$ dex and
$\langle\mathrm{[Fe/H]_{F2}}\rangle = -1.65 \pm 0.11 \pm 0.40$
dex. The first errors take into account the uncertainties of both
periods and amplitudes, whereas the second errors are just the
standard deviation from the mean value. There is clearly insufficient
statistical evidence (given the estimated uncertainties) to conclude
that we are studying two completely different populations, although
the (insignificantly) higher metallicity of the RR Lyrae stars in F1
is in line with the results of M09.  Note however that, as stated by
\citet{Alcock00}, this relation may be affected by evolutionary
effects on the RRab variables and age effects in the stellar
population.

\subsubsection{RR Lyrae Distance Modulus: Luminosity--Metallicity
  Relation and FOBE Method}

Much has been written about the luminosities of horizontal branch (HB)
stars (particularly their $V$-band magnitudes) and their dependence on
metallicity. Recent reviews of the subject include papers by
\citet{Chaboyer99,Cacciari99,Cacciari03}, and \citet{CC03}. In this
last paper, the authors describe the main properties of HB
luminosities as follows:
\begin{itemize}
\item the relation $M_V(RR) = \alpha\mathrm{[Fe/H]} + \beta$
  \citep{Sandage81a,Sandage81b} can be considered linear, as a first
  approximation, with a variation of $\sim0.25$ mag over 1 dex in
  metallicity;
\item at a given metallicity the evolution of low mass stars from the
  HB produces an intrinsic magnitude spread that can reach up to
  $\sim0.5$ mag.  This spread can increase as the metallicity of the
  observed cluster increases \citep{Sandage90};
\item the luminosity--metallicity relation is not strictly linear but
  depends on the HB morphology and is related to the ``second
  parameter'' problem \citep{Caputo00,Demarque00}.
\end{itemize}

RR Lyrae variables remain however excellent distance indicators once
these effects are properly known and taken into account.  We assume a
linear $M_V(RR)$--[Fe/H] relation, the average of several methods:
\begin{equation}
M_V(RR) = (0.23 \pm0.04)\mathrm{[Fe/H]} + (0.93 \pm 0.12)
\end{equation}
\citep[for further details see][]{CC03}. To derive the distance
modulus, we used this equation with observed mean magnitudes of
$\langle V\rangle=25.34\pm0.15$ mag for F1 and $\langle
V\rangle=25.30\pm0.12$ mag for F2 and a metallicity of
$\mathrm{[Fe/H]}=-1.6$ dex for both fields (roughly the mean of our
inferred metallicities in Sec.~5.3.1).  Furthermore, we assumed
different values for the reddening values as derived and discussed in
Section~\ref{sec:RE}:
\begin{description}
\item[(a)] $\mu_0(\mathrm{F1})=24.53\pm0.21$ mag and
  $\mu_0(\mathrm{F2})=24.49\pm0.19$ mag by using $E(B-V)=0.08$ mag for
  both fields, according to the \citet{Schlegel98} map;
\item[(b)] $\mu_0(\mathrm{F1})=24.56\pm0.21$ mag and
  $\mu_0(\mathrm{F2})=24.64\pm0.19$ mag by using $E(B-V)=0.07$ mag and
  $E(B-V)=0.03$ respectively for F1 and F2, according to the
  \citet{Sturch66} method; and
\item[(c)] $\mu_0(\mathrm{F1})=24.72\pm0.21$ mag and
  $\mu_0(\mathrm{F2})=24.59\pm0.19$ mag by using $E(B-V)=0.02$ mag for
  both fields, according to the FOBE method.
\end{description}
The errors are conservative and take into account the scatter in the
mean $\langle V\rangle$ value, metallicity uncertainties (of about
0.10 dex), and the intrinsic uncertainty in the $M_V(RR)$--[Fe/H]
relation. Note that only taking into account for a differential
reddening between the two fields (case b) can one find F1 (the field
closer to M32) to be slightly closer to us than F2.  As shown by S09,
mostly because of the large distance of M31 from us and the relatively
short distance between M31 and M32, these distance moduli do not
provide any further information to disentangle the two
populations. These moduli appear to be in very good agreement with
estimates of both M31 ($\mu_0=24.44\pm0.11$, \citealt{Freedman90};
$\mu_0=24.5\pm0.1$, \citealt{Brown04}; $\mu_0=24.47\pm0.07$,
\citealt{Mcconnachie05}; $\mu_0=24.54 \pm 0.07$, \citealt{Saha06};
$\mu_0=24.46\pm0.11$, S09) and M32 ($\mu_0=24.55\pm0.08$,
\citealt{Tonry01}; $\mu_0=24.39\pm0.08$, \citealt{Jensen03}) distance
moduli and with a recent estimate of the distance to M32 obtained with
our data using the Red Clump method, $\mu_0(\mathrm{F1}) =
24.50\pm0.12$ mag (M09).

\begin{figure}
\plotone{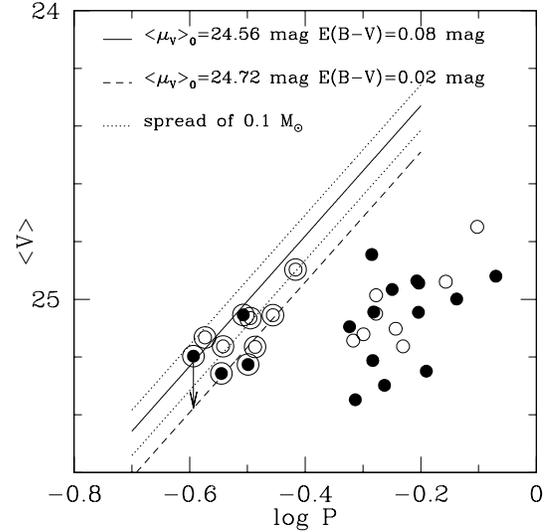}
\caption{RR Lyrae variable stars in the $\langle V\rangle$--$\log P$
  plane, where the visualization of the FOBE method is shown for an
  independent distance modulus (DM) determination. The symbols are the
  same as in Figure~\ref{strip}. The solid line represents the FOBE
  edge, by assuming a stellar mass value of $M=0.7\,M_{\odot}$ and a
  metallicity $\mathrm{[Fe/H]}=-1.6$. With these assumptions we find a
  DM of $\mu_0=24.56$ for a reddening value of 0.08 mag for both
  fields. If we assume the minimum reddening estimate of 0.02 mag as
  we derived (FOBE method, see Section~\ref{sec:RE}), all the stars
  move towards fainter magnitudes an amount indicated by the arrow. In
  this case, we obtain a DM of $\mu_0=24.72$ mag (shown by the dashed
  line). The dotted lines represent the FOBE edges we obtained by
  taking into account a $0.1\,M_{\odot}$ spread in stellar mass around
  the initial assumed value.}
\label{fobe}
\end{figure}

The FOBE method described in \citet{Caputo97} and \citet{Caputo00}
provides an independent distance estimator. It is a graphical method
based on the predicted period--luminosity (PL) relation for pulsators
located along the first-overtone blue edge (FOBE) and seems quite
robust for clusters with significant numbers of RRc variables.  Using
this procedure, one obtains a distribution of the cluster RR Lyrae
stars in the $M_{V}$--$\log P$ plane once a distance modulus has been
assumed. By matching the observed distribution of RRc variables with
the following theoretical relation for the FOBE \citep{Caputo00},
\begin{eqnarray}
M_V(\mathrm{FOBE}) &=& -0.685(\pm0.027) - 2.255 \log
P(\mathrm{FOBE}) \nonumber\\ 
& &\quad - 1.259\log(M/M_{\odot}) + 0.058 \log Z,
\end{eqnarray}
we can obtain an independent estimate of the distance modulus. We
assume the mean metallicities derived above, $\mathrm{[Fe/H]} = -1.6$
for both fields, corresponding to $\log Z = -3.3$, and we take $M =
0.7 M_{\odot}$ from evolutionary HB models for RRc variables, with an
uncertainty of the order of 4\% \citep{Bono03}. The FOBE method then
yields a distance modulus of $\mu_0=24.56\pm0.10$ mag for both fields
assuming $E(B-V)=0.08$ mag.
If we also take into account the minimum reddening value we obtained
in Section~\ref{sec:RE} of 0.02 mag, the distance modulus increases
to $\mu_0=24.72\pm0.10$ mag.

\subsubsection{PLC and PLC-Amplitude Relations}

\begin{figure}
\plotone{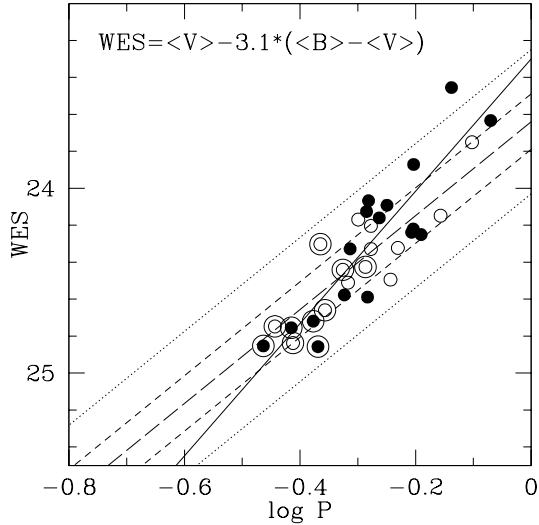}
\caption{RR Lyrae stars in the reddening-free Wesenheit plane:
  $\langle V\rangle-3.1\times(B-V)$ vs.\ $\log P$. The symbols are the
  same as in Figure~\ref{strip}. Solid and dashed lines represent the
  linear fits to F1 and F2 datasets respectively. The slopes of these
  relationships show a small but insignificant difference, as the F1
  relationship is well within the $1\sigma$ and $3\sigma$
  uncertainties of the F2 relation (thin dashed and dotted lines,
  respectively).}
\label{wes}
\end{figure}

\begin{figure}
\plotone{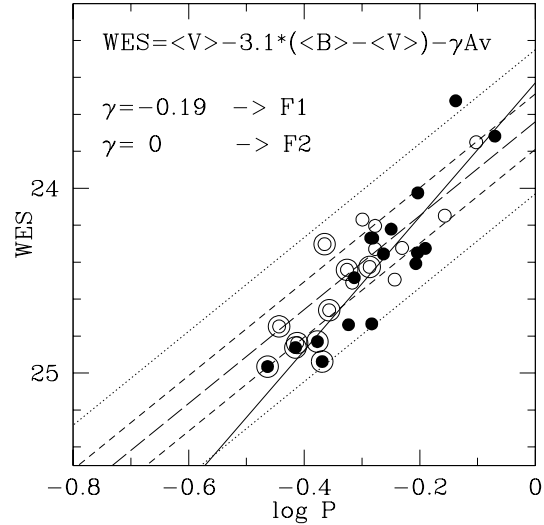}
\caption{RR Lyrae stars in the reddening-free Wesenheit plane, taking
  into account the $V$-band amplitudes: $\langle V\rangle-3.1 \times
  (<B>-<V>)-\gamma A_V$ vs $\log P$. The symbols are the same as in
  Figure~\ref{strip}. Solid and dashed solid lines represent the fits
  to our data obtained with $+0.01\gamma$ and $-0.29\gamma$ values for
  F1 and F2, respectively.  There is again an insignificant difference
  between the two fits, as shown in the $1\sigma$ and $3\sigma$
  uncertainties for the F2 relation (thin dashed and dotted lines,
  respectively). }
\label{wes-av}
\end{figure}

Our final attempt to disentangle the two populations is to consider at
the same time all the information on the RR Lyrae variables we have
obtained (including magnitudes, amplitudes, colors and periods) that
reflect the individual properties of these stars.  In
Figures~\ref{wes} and \ref{wes-av} we show our detected RR Lyrae stars
in the reddening-free Wesenheit plane, $\langle
V\rangle-R_V(<B>-<V>)-\gamma A_V$ vs period, without ($\gamma=0$) or
with ($\gamma\neq0$) taking into account a dependence on the stellar
amplitudes, respectively. Because our sample of RR Lyrae variables is
very small, we ``fundamentalized'' the first-overtone (FO) pulsators
using the relation $\log P_{F}=\log P_{FO} + 0.13$, where $\log P_{F}$
is the fundamentalized period \citep{vAB73}. In Figure~\ref{wes} we
see that both RR Lyrae groups are located in the same region; the
solid and dashed lines represent the best linear fit across the data
points for F1 and F2, respectively. As we can see only the slope of
the relation changes, but this slope difference is statistically
insignificant. The thin dashed and dotted lines represent the
dispersion around the relation for F2, for $1\sigma$ and $3\sigma$
respectively.  In Figure~\ref{wes-av} we see that the periods of RR
Lyrae stars in F2 do not show any dependence on their own amplitudes,
whereas the periods of RR Lyrae stars in F1 show a mild dependence on
their pulsation amplitudes of $\gamma=-0.19$ dex.  However, the two
relationships are in statistical agreement with each other, suggesting
that the RR Lyrae properties of the two groups are very similar.

\smallskip
We summarize this section by stating that the properties of RR Lyrae
variables in our two fields, F1 and F2, are statistically inseparable.
The only difference we find between the two fields is in the color of
the FRE, where a population of F1 RRab variables are redder than the
F2 RRab variables.  We suggest that these stars may in fact belong to
the M32 RR Lyrae population, but the small number of detected RR Lyrae
stars prevent us from making a definitive claim as to the true cause
of this color difference.

\section{What Have We Learned About M32, and What Do We Still Not Know?}
\label{sec:conclusions}

In this paper we present ACS/HRC observations of fields near M32 (and
supplemental ACS/WFC observations) to search for RR Lyrae variable
stars. The detection of RR Lyrae variable stars represents the only
way to obtain information about the presence of an ancient, metal-poor
population ($\ga10$ Gyr and $\mathrm{[Fe/H]}\la-1$ dex) in M32 from
optical data not deep enough to detect the oldest MSTO stars.

We have detected 17 RR Lyrae variable stars in F1, our field closest
to M32, and 14 RR Lyrae variable stars in F2, our background field in
M31's disk.  We can only claim to have detected an upper limit of 6 RR
Lyrae stars belonging to M32 in F1 based on these two fields alone.
We can better constrain the M32 RR Lyrae population by extending our
analysis to our ACS/WFC parallel fields F3 and F4 and by assuming that
the M31 surface brightness is constant over these fields.  With this
extension, we can claim to have detected $7_{-3}^{+4}$ RR Lyrae
variable stars belonging to M32 in F1, and therefore we claim that
\emph{M32 has an ancient population}.  The implied specific frequency
of RR Lyrae stars in M32 is $\mathrm{S_{RR}}\approx6.5$, with a 68\%
confidence interval of $3.6\leq\mathrm{S_{RR}}\leq10$, similar to
typical intermediate-metallicity ($\mathrm{[Fe/H]}\sim-1.5$ dex)
Galactic globular clusters.  In making these estimates, note that we
have assumed that the metal-poor population constitutes 11\% of the
total luminosity of M32 in F1 following the metallicity distribution
function presented in M09 and further that this population has an age
of 10 Gyr.

We have also used the background M31 population to infer a specific
frequency of RR Lyrae variables in M31's disk,
$\mathrm{S_{RR}}\approx18$, that is higher than the value found by
\citet{Brown04} in M31's halo, $\mathrm{S_{RR}}\approx11$.  This may
reflect merely small fluctuations in the horizontal-branch morphology
with position in M31 or it may reflect some property of M31's disk; it
is difficult to state conclusively the cause without a better
understanding of how the horizontal branch is populated (and therefore
how mass is lost on the first-ascent giant branch).

Even though we claim to have detected \emph{bona fide} M32 RR Lyrae
variable stars in F1, the pulsational properties stars in fields F1
and F2 present nearly no significant differences. They have
indistinguishable mean $V$ magnitudes [$\langle V\rangle=25.34\pm0.15$
  mag and $\mu_0=24.53\pm0.21$ mag for F1; $\langle
  V\rangle=25.30\pm0.12$ mag and $\mu_0=24.49\pm0.19$ mag for F2 by
  assuming E(B-V)=0.08 mag and $\mathrm{[Fe/H]}=-1.6$ dex for both
  fields], the same mean periods ($\langle P_{ab}\rangle=0.59\pm0.11$
d), and the same distribution in the Bailey ($V$-band amplitude--log
Period) diagram, and insignificantly different ratios of RRc to RRab
types ($N_c/N_{ab}=0.30_{-0.30}^{+0.37}$ and
$N_c/N_{ab}=0.75_{-0.42}^{+1.25}$ respectively). By using the relation
between the metallicity [Fe/H] of each RRab variable, its period
$P_{ab}$, and its $V$-band amplitude $A_V$, we find mean values of
$\langle\mathrm{[Fe/H]}\rangle_{\mathrm{F1}}=-1.52\pm0.10$ dex and
$\langle\mathrm{[Fe/H]}\rangle_{\mathrm{F2}}=-1.65\pm0.11$ dex. We
stress here that these metallicities, as well as those found by S09
($\mathrm{[Fe/H]}=-1.77\pm0.06$ from nearly 700 stars), are far away
from the ``solar-like'' metallicity used by \citet{Brown00} to
interpret the UV excess they found in the core of M32 as a hot
horizontal branch. However, we do find a population of F1 RRab
variables that are redder than F2 RRab variables.  We suggest that
these stars may in fact belong to the M32 RR Lyrae population, but the
small-number statistics imposed by the small ACS/HRC area makes a
definitive claim difficult or even impossible.

We have detected M32's long sought-after ancient, metal-poor
population through its RR Lyrae population.  But we find that this RR
Lyrae population is nearly indistinguishable in its mean pulsational
properties from M31, and for that matter, its specific frequency.
Does this imply that M32 and M31 formed, evolved, and (likely)
interacted in such a way that their ancient, metal-poor populations
share some commonality?  Or is this just a coincidence?  We hope that
further exploration of M32's RR Lyrae population, from fields closer
to the center of M32, and high-resolution spectroscopy of individual
stars, revealing their detailed chemical compositions and therefore
details about their formation histories, in M32 and the disk of M31
will eventually become available to help answer these questions.

\acknowledgments

Support for program GO-10572 was provided by NASA through a grant from
the Space Telescope Science Institute, which is operated by the
Association of Universities for Research in Astronomy, Inc., under
NASA contract NAS 5-26555.  We really thank E. Bernard, M. Monelli,
R. Ramos Contreras, G. Clementini and M. Irwin for interesting
suggestions and comments about this work and A. Sarajedini,
L. Koopmans, and S. Vegetti for helpful discussions.

{\it Facility:} \facility{HST (ACS)}

\bibliographystyle{apj}
\bibliography{m32}

\begin{thebibliography}{77}
\expandafter\ifx\csname natexlab\endcsname\relax\def\natexlab#1{#1}\fi

\bibitem[{{Alcock} {et~al.}(2000)}]{Alcock00}
{Alcock}, C., {et~al.} 2000, \aj, 119, 2194

\bibitem[{{Alonso-Garc{\'{\i}}a} {et~al.}(2004){Alonso-Garc{\'{\i}}a}, {Mateo},
  \& {Worthey}}]{AlonsoGarcia04}
{Alonso-Garc{\'{\i}}a}, J., {Mateo}, M., \& {Worthey}, G. 2004, \aj, 127, 868

\bibitem[{{Baker} \& {Kippenhahn}(1962)}]{BK62}
{Baker}, N., \& {Kippenhahn}, R. 1962, Zeitschrift fur Astrophysik, 54, 114

\bibitem[{{Bernard} {et~al.}(2009)}]{Bernard09}
{Bernard}, E.~J., {et~al.} 2009, \apj, 699, 1742

\bibitem[{{Blanco}(1992)}]{Blanco92}
{Blanco}, V.~M. 1992, \aj, 104, 734

\bibitem[{{Bono} {et~al.}(2003){Bono}, {Caputo}, {Castellani}, {Marconi},
  {Storm}, \& {Degl'Innocenti}}]{Bono03}
{Bono}, G., {Caputo}, F., {Castellani}, V., {Marconi}, M., {Storm}, J., \&
  {Degl'Innocenti}, S. 2003, \mnras, 344, 1097

\bibitem[{{Bono} \& {Stellingwerf}(1994)}]{BS94}
{Bono}, G., \& {Stellingwerf}, R.~F. 1994, \apjs, 93, 233

\bibitem[{{Brown} {et~al.}(2000){Brown}, {Bowers}, {Kimble}, {Sweigart}, \&
  {Ferguson}}]{Brown00}
{Brown}, T.~M., {Bowers}, C.~W., {Kimble}, R.~A., {Sweigart}, A.~V., \&
  {Ferguson}, H.~C. 2000, \apj, 532, 308

\bibitem[{{Brown} {et~al.}(2004){Brown}, {Ferguson}, {Smith}, {Kimble},
  {Sweigart}, {Renzini}, \& {Rich}}]{Brown04}
{Brown}, T.~M., {Ferguson}, H.~C., {Smith}, E., {Kimble}, R.~A., {Sweigart},
  A.~V., {Renzini}, A., \& {Rich}, R.~M. 2004, \aj, 127, 2738

\bibitem[{{Burstein} \& {Heiles}(1978)}]{BH78}
{Burstein}, D., \& {Heiles}, C. 1978, \apj, 225, 40

\bibitem[{{Burstein} \& {Heiles}(1982)}]{BH82}
---. 1982, \aj, 87, 1165

\bibitem[{{Cacciari}(1999)}]{Cacciari99}
{Cacciari}, C. 1999, in Astronomical Society of the Pacific Conference Series,
  Vol. 167, Harmonizing Cosmic Distance Scales in a Post-HIPPARCOS Era, ed.
  D.~{Egret} \& A.~{Heck}, 140

\bibitem[{{Cacciari}(2003)}]{Cacciari03}
{Cacciari}, C. 2003, in Astronomical Society of the Pacific Conference Series,
  Vol. 296, New Horizons in Globular Cluster Astronomy, ed. G.~{Piotto}
  {et~al.}, 329

\bibitem[{{Cacciari} \& {Clementini}(2003)}]{CC03}
{Cacciari}, C., \& {Clementini}, G. 2003, in Lecture Notes in Physics, Berlin
  Springer Verlag, Vol. 635, Stellar Candles for the Extragalactic Distance
  Scale, ed. D.~{Alloin} \& W.~{Gieren}, 105

\bibitem[{{Caputo}(1997)}]{Caputo97}
{Caputo}, F. 1997, \mnras, 284, 994

\bibitem[{{Caputo} {et~al.}(2000){Caputo}, {Castellani}, {Marconi}, \&
  {Ripepi}}]{Caputo00}
{Caputo}, F., {Castellani}, V., {Marconi}, M., \& {Ripepi}, V. 2000, \mnras,
  316, 819

\bibitem[{{Catelan}(2006)}]{Catelan06}
{Catelan}, M. 2006, in Revista Mexicana de Astronomia y Astrofisica, vol. 27,
  Vol.~26, Revista Mexicana de Astronomia y Astrofisica Conference Series, 93

\bibitem[{{Chaboyer}(1999)}]{Chaboyer99}
{Chaboyer}, B. 1999, in Astrophysics and Space Science Library, Vol. 237,
  Post-Hipparcos cosmic candles, ed. A.~{Heck} \& F.~{Caputo}, 111

\bibitem[{{Choi} {et~al.}(2002){Choi}, {Guhathakurta}, \& {Johnston}}]{choi}
{Choi}, P.~I., {Guhathakurta}, P., \& {Johnston}, K.~V. 2002, \aj, 124, 310

\bibitem[{{Clement}(2000)}]{Clement00}
{Clement}, C.~M. 2000, in Astronomical Society of the Pacific Conference
  Series, Vol. 203, IAU Colloq. 176: The Impact of Large-Scale Surveys on
  Pulsating Star Research, ed. L.~{Szabados} \& D.~{Kurtz}, 266

\bibitem[{{Clement} {et~al.}(2001)}]{Clement01}
{Clement}, C.~M., {et~al.} 2001, \aj, 122, 2587

\bibitem[{{Clementini} {et~al.}(2003){Clementini}, {Gratton}, {Bragaglia},
  {Carretta}, {Di Fabrizio}, \& {Maio}}]{Clementini03}
{Clementini}, G., {Gratton}, R., {Bragaglia}, A., {Carretta}, E., {Di
  Fabrizio}, L., \& {Maio}, M. 2003, \aj, 125, 1309

\bibitem[{{Clementini} {et~al.}(2009)}]{Clementini09} 
{Clementini}, G, {Contreras}, R., {Federici}, L., {Cacciari}, C., {Merighi}, R.,  {Smith}, H., {Catelan}, M., {Fusi Pecci}, F., {Marconi}, M.,  {Kinemuchi}, K.,  {Pritzl}, B., 2009, \apj, in press, arXiv:0909.3303v1

\bibitem[{{Clementini} {et~al.}(2000)}]{Clementini00}
{Clementini}, G., {et~al.} 2000, \aj, 120, 2054

\bibitem[{{Coelho} {et~al.}(2009){Coelho}, {Mendes de Oliveira}, \&
  {Fernandes}}]{Coelho09}
{Coelho}, P., {Mendes de Oliveira}, C., \& {Fernandes}, R.~C. 2009, \mnras,
  396, 624


\bibitem[{{Cox}(1963)}]{Cox63}
{Cox}, J.~P. 1963, \apj, 138, 487

\bibitem[{{Demarque} {et~al.}(2000){Demarque}, {Zinn}, {Lee}, \&
  {Yi}}]{Demarque00}
{Demarque}, P., {Zinn}, R., {Lee}, Y.-W., \& {Yi}, S. 2000, \aj, 119, 1398

\bibitem[{{Dolphin}(2000)}]{Dolphin00}
{Dolphin}, A.~E. 2000, \pasp, 112, 1383

\bibitem[{{Eddington}(1926)}]{Eddington26}
{Eddington}, A.~S. 1926, The Observatory, 49, 88

\bibitem[{{Fiorentino} {et~al.}(2002){Fiorentino}, {Caputo}, {Marconi}, \&
  {Musella}}]{Fiorentino02}
{Fiorentino}, G., {Caputo}, F., {Marconi}, M., \& {Musella}, I. 2002, \apj,
  576, 402

\bibitem[{{Freedman} \& {Madore}(1990)}]{Freedman90}
{Freedman}, W.~L., \& {Madore}, B.~F. 1990, \apj, 365, 186

\bibitem[{{Galleti} {et~al.}(2006){Galleti}, {Federici}, {Bellazzini},
  {Buzzoni}, \& {Pecci}}]{Galleti06}
{Galleti}, S., {Federici}, L., {Bellazzini}, M., {Buzzoni}, A., \& {Pecci},
  F.~F. 2006, \apjl, 650, L107

\bibitem[{{Gonz{\'a}lez}(1993)}]{G93}
{Gonz{\'a}lez}, J.~J. 1993, PhD thesis, University of California, Santa Cruz

\bibitem[{{Harris}(1996)}]{Harris96}
{Harris}, W.~E. 1996, \aj, 112, 1487

\bibitem[{{Holtzman} {et~al.}(1995){Holtzman}, {Burrows}, {Casertano},
  {Hester}, {Trauger}, {Watson}, \& {Worthey}}]{Holtzman95}
{Holtzman}, J.~A., {Burrows}, C.~J., {Casertano}, S., {Hester}, J.~J.,
  {Trauger}, J.~T., {Watson}, A.~M., \& {Worthey}, G. 1995, \pasp, 107, 1065

\bibitem[{{Jensen} {et~al.}(2003){Jensen}, {Tonry}, {Barris}, {Thompson},
  {Liu}, {Rieke}, {Ajhar}, \& {Blakeslee}}]{Jensen03}
{Jensen}, J.~B., {Tonry}, J.~L., {Barris}, B.~J., {Thompson}, R.~I., {Liu},
  M.~C., {Rieke}, M.~J., {Ajhar}, E.~A., \& {Blakeslee}, J.~P. 2003, \apj, 583,
  712

\bibitem[{{King} \& {Cox}(1968)}]{KC68}
{King}, D.~S., \& {Cox}, J.~P. 1968, \pasp, 80, 365

\bibitem[{{Kormendy} {et~al.}(2009){Kormendy}, {Fisher}, {Cornell}, \&
  {Bender}}]{kfc}
{Kormendy}, J., {Fisher}, D.~B., {Cornell}, M.~E., \& {Bender}, R. 2009, \apjs,
  182, 216

\bibitem[{{Lafler} \& {Kinman}(1965)}]{LK65}
{Lafler}, J., \& {Kinman}, T.~D. 1965, \apjs, 11, 216

\bibitem[{{Lauer}(1999)}]{l99}
{Lauer}, T.~R. 1999, \pasp, 111, 227

\bibitem[{{McConnachie} {et~al.}(2005){McConnachie}, {Irwin}, {Ferguson},
  {Ibata}, {Lewis}, \& {Tanvir}}]{Mcconnachie05}
{McConnachie}, A.~W., {Irwin}, M.~J., {Ferguson}, A.~M.~N., {Ibata}, R.~A.,
  {Lewis}, G.~F., \& {Tanvir}, N. 2005, \mnras, 356, 979

\bibitem[{{McDonald}(1977)}]{McDonald77}
{McDonald}, L.~H. 1977, PhD thesis, University of California, Santa Cruz

\bibitem[{{Monachesi} {et~al.}(2009){Monachesi}, {Trager}, {Lauer}, {Freedman},
  {Dressler}, {Mighell}, \& {Grillmair}}]{M09}
{Monachesi}, A., {Trager}, S.~C., {Lauer}, T.~R., {Freedman}, W.~L.,
  {Dressler}, A., {Mighell}, K.~J., \& {Grillmair}, C.~J. 2009, \apj, submitted

\bibitem[{{O'Connell}(1980)}]{OConnell80}
{O'Connell}, R.~W. 1980, \apj, 236, 430

\bibitem[{{Oosterhoff}(1939)}]{Oosterhoff39}
{Oosterhoff}, P.~T. 1939, The Observatory, 62, 104

\bibitem[{{Pritzl} {et~al.}(2002){Pritzl}, {Armandroff}, {Jacoby}, \& {Da
  Costa}}]{Pritzl02}
{Pritzl}, B.~J., {Armandroff}, T.~E., {Jacoby}, G.~H., \& {Da Costa}, G.~S.
  2002, \aj, 124, 1464

\bibitem[{{Pritzl} {et~al.}(2004){Pritzl}, {Armandroff}, {Jacoby}, \& {Da
  Costa}}]{Pritzl04}
---. 2004, \aj, 127, 318

\bibitem[{{Rose}(1985)}]{Rose85}
{Rose}, J.~A. 1985, \aj, 90, 1927

\bibitem[{{Rose}(1994)}]{Rose94}
---. 1994, \aj, 107, 206

\bibitem[{{Saha} \& {Hoessel}(1990)}]{SH90}
{Saha}, A., \& {Hoessel}, J.~G. 1990, \aj, 99, 97

\bibitem[{{Saha} {et~al.}(1986){Saha}, {Monet}, \& {Seitzer}}]{Saha86}
{Saha}, A., {Monet}, D.~G., \& {Seitzer}, P. 1986, \aj, 92, 302

\bibitem[{{Saha} {et~al.}(2006){Saha}, {Thim}, {Tammann}, {Reindl}, \&
  {Sandage}}]{Saha06}
{Saha}, A., {Thim}, F., {Tammann}, G.~A., {Reindl}, B., \& {Sandage}, A. 2006,
  \apjs, 165, 108

\bibitem[{{Salaris} \& {Cassisi}(2005)}]{SCbook}
{Salaris}, M., \& {Cassisi}, S. 2005, {Evolution of Stars and Stellar
  Populations} (Chichester: John Wiley \& Sons, Ltd)

\bibitem[{{Sandage}(1981{\natexlab{a}})}]{Sandage81a}
{Sandage}, A. 1981{\natexlab{a}}, \apjl, 244, L23

\bibitem[{{Sandage}(1981{\natexlab{b}})}]{Sandage81b}
---. 1981{\natexlab{b}}, \apj, 248, 161

\bibitem[{{Sandage}(1990)}]{Sandage90}
---. 1990, \apj, 350, 603

\bibitem[{{Sarajedini} {et~al.}(2009){Sarajedini}, {Mancone}, {Lauer},
  {Dressler}, {Freedman}, {Trager}, {Grillmair}, \& {Mighell}}]{s09}
{Sarajedini}, A., {Mancone}, C.~L., {Lauer}, T.~R., {Dressler}, A., {Freedman},
  W., {Trager}, S.~C., {Grillmair}, C., \& {Mighell}, K.~J. 2009, \aj, 138, 184

\bibitem[{{Schlegel} {et~al.}(1998){Schlegel}, {Finkbeiner}, \&
  {Davis}}]{Schlegel98}
{Schlegel}, D.~J., {Finkbeiner}, D.~P., \& {Davis}, M. 1998, \apj, 500, 525

\bibitem[{{Sirianni} {et~al.}(2005)}]{Sirianni05}
{Sirianni}, M., {et~al.} 2005, \pasp, 117, 1049

\bibitem[{{Sivia} \& {Skilling}(2006)}]{Sivia06}
{Sivia}, D.~S., \& {Skilling}, J. 2006, {Data Analysis: A Bayesian Tutorial}
  (2nd ed.; Oxford, UK: Oxford University Press)

\bibitem[{{Smith}(1995)}]{Smith95}
{Smith}, H.~A. 1995, {RR Lyrae stars} (Cambridge Astrophysics Series,
  Cambridge, New York: Cambridge University Press)

\bibitem[{{Stellingwerf}(1978)}]{Stellingwerf78}
{Stellingwerf}, R.~F. 1978, \apj, 224, 953

\bibitem[{{Stetson}(1987)}]{Stetson87}
{Stetson}, P.~B. 1987, \pasp, 99, 191

\bibitem[{{Stetson}(1994)}]{Stetson94}
---. 1994, \pasp, 106, 250

\bibitem[{{Sturch}(1966)}]{Sturch66}
{Sturch}, C. 1966, \apj, 143, 774

\bibitem[{{Suntzeff} {et~al.}(1991){Suntzeff}, {Kinman}, \&
  {Kraft}}]{Suntzeff91}
{Suntzeff}, N.~B., {Kinman}, T.~D., \& {Kraft}, R.~P. 1991, \apj, 367, 528

\bibitem[{{Szczygie{\l}} {et~al.}(2009){Szczygie{\l}}, {Pojma{\'n}ski}, \&
  {Pilecki}}]{SPP09}
{Szczygie{\l}}, D.~M., {Pojma{\'n}ski}, G., \& {Pilecki}, B. 2009, Acta
  Astronomica, 59, 137

\bibitem[{{Thomas} {et~al.}(2005){Thomas}, {Maraston}, {Bender}, \& {Mendes de
  Oliveira}}]{TMBO05}
{Thomas}, D., {Maraston}, C., {Bender}, R., \& {Mendes de Oliveira}, C. 2005,
  ApJ, 621, 673

\bibitem[{{Tonry} {et~al.}(2001){Tonry}, {Dressler}, {Blakeslee}, {Ajhar},
  {Fletcher}, {Luppino}, {Metzger}, \& {Moore}}]{Tonry01}
{Tonry}, J.~L., {Dressler}, A., {Blakeslee}, J.~P., {Ajhar}, E.~A., {Fletcher},
  A.~B., {Luppino}, G.~A., {Metzger}, M.~R., \& {Moore}, C.~B. 2001, \apj, 546,
  681

\bibitem[{{Trager} {et~al.}(2000){Trager}, {Faber}, {Worthey}, \&
  {Gonz{\'a}lez}}]{T00b}
{Trager}, S.~C., {Faber}, S.~M., {Worthey}, G., \& {Gonz{\'a}lez}, J.~J. 2000,
  AJ, 120, 165

\bibitem[{{van Albada} \& {Baker}(1973)}]{vAB73}
{van Albada}, T.~S., \& {Baker}, N. 1973, \apj, 185, 477

\bibitem[{{Walker}(1990)}]{Walker90}
{Walker}, A.~R. 1990, \aj, 100, 1532

\bibitem[{{Walker}(1992)}]{Walker92}
---. 1992, \aj, 104, 1395

\bibitem[{{Walker}(1998)}]{Walker98}
---. 1998, \aj, 116, 220

\bibitem[{{Zhevakin}(1953)}]{Zhevakin53}
{Zhevakin}, S.~A. 1953, \azh, 30, 161

\bibitem[{{Zhevakin}(1959)}]{Zhevakin59}
---. 1959, Soviet Astronomy, 3, 913

\bibitem[{{Zinn} \& {West}(1984)}]{ZW84}
{Zinn}, R., \& {West}, M.~J. 1984, \apjs, 55, 45

\end{thebibliography}

\end{document}